\documentstyle[aps,epsfig,preprint]{revtex}
\tighten

\newcommand{\identity}{{1\hspace{-0.6ex}{\mathrm l}}}
\newcommand{\sub}[1]{_{\mbox{\scriptsize\unboldmath$#1$}}}
\newcommand{\super}[1]{^{\mbox{\scriptsize\unboldmath$#1$}}}
\newcommand{\itbf}[1]{\mbox{\boldmath$#1$}}
\newcommand{\smallitbf}[1]{\mbox{\scriptsize\boldmath$#1$}}
\newcommand{\sfrac}[2]{\mbox{$\frac{#1}{#2}$}}
\newcommand{\Q}{{\bf Q}_{\smallitbf{\hat{\sigma}}}}
\newcommand{\euler}{C}

\begin{document}

\title{
Front propagation techniques to calculate the largest Lyapunov
exponent of dilute hard disk gases
}

\author{
R.\ van Zon}
\address{Chemical Physics Theory Group, Department of Chemistry,\\
University of Toronto,
Toronto, Ontario, Canada M5S 3H6
\\
and \\
Institute
for Theoretical
Physics, Utrecht University,
\\
Leuvenlaan 4, 3584 CE Utrecht, The Netherlands}
\author{Henk van Beijeren}
\address{
Institute for Theoretical Physics, University of Utrecht,
\\
Leuvenlaan 4, 3584 CE Utrecht, The Netherlands
\\
and\\
Center for Nonlinear Phenomena and Complex Systems, Universit\'e
Libre de Bruxelles, \\
Campus Plaine, Code Postal 231, 1050 Brussels, Belgium }

\date{\today}

\maketitle

\begin{abstract}
A kinetic approach is adopted to describe the exponential growth of a
small deviation of the initial phase space point, measured by the
largest Lyapunov exponent, for a dilute system of hard disks, both in
equilibrium and in a uniform shear flow. We derive a generalized
Boltzmann equation for an extended one-particle distribution that
includes deviations from the reference phase space point. The equation
is valid for very low densities $n$, and requires an unusual expansion
in powers of $1/|\ln n|$.  It reproduces and extends results from the
earlier, more heuristic clock model and may be interpreted as
describing a front propagating into an unstable state.  The asymptotic
speed of propagation of the front is proportional to the largest
Lyapunov exponent of the system.  Its value may be found by applying
the standard front speed selection mechanism for pulled fronts to the
case at hand.  For the equilibrium case, an explicit expression for
the largest Lyapunov exponent is given and for sheared systems we give
explicit expressions that may be evaluated numerically to obtain the
shear rate dependence of the largest Lyapunov exponent.
\if preprintsty \else \vspace{3mm} \fi
\end{abstract}

\section{Introduction}

Often, familiar models from statistical mechanics exhibit the strong
dependence on the initial phase space point that we know as
chaos\cite{Ott}. Examples are the Lorentz gas, the hard disk and the
hard sphere gas\cite{szasz}. We call a system {\em chaotic} if the
growth of a deviation from a reference trajectory in phase space is
exponential, $\propto \exp(\lambda^+t)$, in the limit that the initial
deviation becomes infinitesimally small. $\lambda^+$ is called the
largest Lyapunov exponent.

The systems we discuss in this paper, consist of $N$ hard disks with
equal mass $m$ and equal diameters $a$, in a two dimensional volume
$V$. The density is $n=N/V$, and a (dimensionless) reduced density is
defined as $\tilde{n}=na^2$. The position and velocity of disk $l$ are
denoted by $\itbf{r\sub{l}}$ and $\itbf{v\sub{l}}$ respectively.  We
define a temperature in equilibrium by $N k_BT = \langle \sum_{l=1}^N
\sfrac{1}{2} m |\itbf{v\sub{l}}|^2 \rangle$, where the brackets denote
an ensemble average.  For non-equilibrium stationary states, a
generalization of this is used for the temperature.  A typical
velocity $v_0$ is defined as $v_0 =\sqrt{k_BT/m}$.

Considering a point in phase space $\{\itbf{r\sub{h}},
\itbf{v\sub{h}}\}$ ($h=1\ldots N$) and an adjacent point
$\{\itbf{r\sub{h}\super{*}}, \itbf{v\sub{h}\super{*}}\}$ with
$\itbf{r\sub{h}\super{*}}=\itbf{r\sub h}+\itbf{\delta r\sub{h}}$
and $\itbf{v\sub{h}\super{*}}=\itbf{v\sub{h}}+\itbf{\delta v\sub{h}}$,
one may obtain the largest Lyapunov exponent by studying the
exponential growth of the deviations $\itbf{\delta r\sub{h}}$ and
$\itbf{\delta v\sub{h}}$.  Other Lyapunov exponents exist, measuring
growth rates of deviations in carefully selected different directions.
Typical deviations have a component in the most rapidly expanding
direction, so they grow with the largest Lyapunov exponent.

Of this chaotic behavior, little is seen on a macroscopic level,
especially when we consider systems in or near equilibrium, but
surprisingly, from numerical simulations, one found relations between
Lyapunov exponents in stationary non-equilibrium systems, and linear
transport coefficients\cite{EvansCohenMorriss,Cohentransport,Dellago2}.
Even if the generality of such relations can be questioned\cite{Klages},
it is still interesting to consider models in which they hold. Among
these are particle systems subject to a velocity independent external
force and kept at a constant kinetic energy by means of a Gaussian
thermostat\cite{Leb,DettmanMorriss}.

For the last six years, efforts have been made to get an analytic
grasp on the Lyapunov exponents. For the Lorentz gas at low density in
two and three dimensions, Van Beijeren, Dorfman and co-workers have
set up a kinetic theory in which all Lyapunov exponents could be
calculated and relations with transport coefficients could be
verified\cite{BeijerenDorfman,Beijerenetal,Dorfman2,KSentropy,Latz,BeijerenLatzDorfman,DorfmanAltenberg,dorfmanbook,panja,kruis}.
The hard disk and the hard sphere gas, which were next in line for
kinetic investigation, proved harder to deal with. It is possible to
obtain estimates for the largest Lyapunov exponent in equilibrium for
these systems, based on heuristic effective dynamics of
deviations\cite{myself,leiden,inszasz}. Some results also have been
obtained for the sum of all positive Lyapunov exponents, the
KS-entropy\cite{inszasz,KSentropy2}.

In these approaches, one uses the linear dependence of the deviations
after a collision on their values before, together with the linear
growth of position deviations during the times between collisions.
{}From this, it was argued in Refs.~\cite{myself,inszasz} that {\em at
low densities}, it is enough to look at the {\em clock value}
\begin{equation}
	k_l = \frac{\ln(|\itbf{\delta v\sub{l}}|
		/\delta	v_0)}{|\ln\tilde{n}|},
\label{eq1}
\end{equation}
where $\delta v_0$ is an arbitrary unit of infinitesimal velocity.
These clock values approximately change in a collision between
particle $i$ and~$j$ according to
\begin{equation}
	k_i' = k_j' = \mbox{max}(k_i,k_j) + 1
			+{\cal O}\left(1/\ln\tilde{n}\right),
\label{eq2}
\end{equation}
which is valid to leading order in $\tilde{n}$.  The clock values $k$
could therefore be restricted to integer values.
In a chaotic system, each deviation ${\itbf{\delta v}}_l$ is expected
to grow exponentially and the clock value $k_l$ to grow linearly with
time. Because not all deviations of the particles are identical, there
is a distribution of clock values. The dynamics described by
Eq.~(\ref{eq2}) tends to bring clock values far below average at any
given time back towards the mean. As a result of this the distribution
of all clock values with respect to the instantaneous average for long
times approaches a stationary mean profile, about which the actual
distribution keeps fluctuating. It then follows that the average clock
value increases by 1 plus half the average difference between clock
values per unit time (the mean free time between collisions for a
single particle). In an abstract sense, this is equivalent to the
situation of a propagating front, where a stable phase propagates into
an unstable phase.  The propagation occurs along a $k$-axis. One of
the phases for a given ``position'' $k$ is to have the clock
distribution concentrated to its left, i.e., around lower clock
values. This distribution will shift to the right and go beyond
$k$. So this phase is unstable. The stable phase for a ``position''
$k$ is to have the distribution concentrated to its right. On a
technical note, it turns out that this problem falls into the class of
pulled fronts\cite{wimute}, which is fortunate as it is known how to
obtain the front propagation speed $w$ for such systems.  Using those
techiques, for long times the average clock values
was found to behave as
\[
   \bar{k}= N^{-1}\sum_{l=1}^N k_l(t) = k_0 + w\bar{\nu} t,
\]
with $\bar{\nu}$ the average collision frequency and $w$ a constant of
order $\tilde{n}^0$.  As a consequence of Eq.~(\ref{eq1}) the largest
Lyapunov exponent is
\begin{equation}
	\lambda^+ = - w \bar{\nu} \ln\tilde{n}.
\label{eq3}
\end{equation}
to leading order in the density.
In \cite{inszasz} this analysis was refined further. It was recognized
that the distribution of clock values depends explicitly on the speed
of the particles and the equations describing its time evolution were
adjusted accordingly. Nonetheless the pulled front analogy remains
fully valid.

The purpose of this paper is to give a firmer basis to the heuristic
arguments leading to
equation (\ref{eq3}),
and to extend the methods so as to be applicable to higher densities
and to non-equilibrium cases.

A typical example of a non-equilibrium case is the hard disk gas in a
uniform shearing state, described by the so-called SLLOD
equations\cite{Evans}.  The gas is confined between two infinite
(one-dimensional) parallel plates a distance $2L$ apart, moving in
opposite directions with velocities $ \pm \gamma L$ (see
Fig.~\ref{fig:shear}).  We look at the limit of large $L$ and infinite
extension in the $x$-direction, with fixed shear rate $\gamma$ and
density $n=N/V$.  For small shear rate, a linear velocity profile
develops in the system, so that the fluid velocity at $y$ is
$\itbf{u}=\gamma y\itbf{\hat{x}}$.  We remark that, although we want
the volume $V$ to be macroscopic, $L$ should not be too large,
otherwise the Reynolds number would become so high that the system
becomes turbulent and the assumed linear velocity profile breaks down.

The paper is set up as follows.  In Sec.~\ref{sec:smalldev} we discuss
the dynamics of deviations. In Sec.~\ref{sec:shear} we consider the
velocity distribution function and the distribution of the duration of
intercollisional
flights for the hard disk gas under shear. In Sec.~\ref{sec:EBE} we
set up a generalized Boltzmann equation for a one particle
distribution function that includes the deviations, and expand that in
powers of $1/\ln \tilde{n}$.  In Sec.~\ref{sec:pulled}, this equation
is reinterpreted as describing a propagating front. The largest
Lyapunov exponent is proportional to the front velocity $w$, which we
can determine using a standard method for pulled fronts\cite{wimute}.
The perturbation expansion in the density is further developed in
Sec.~\ref{sec:pertdens}.  In Sec.~\ref{sec:equilibrium} we calculate
the first two terms in the density expansion of the largest Lyapunov
exponent for the two-dimensional hard disk gas in equilibrium.  The
results are formally extended to the shear case in
Sec.~\ref{sec:shear2}. We conclude with a discussion in
Sec.~\ref{sec:conclusions}.

\section{Chaos in hard disk gases}
\label{sec:smalldev}

The dynamics of the hard disk system is defined as follows. When
disks~$i$ and~$j$ hit each other, they collide elastically. We define
$\itbf{r\sub{ij}}=\itbf{r\sub{i}}-\itbf{r\sub{j}}$ and
$\itbf{v\sub{ij}}=\itbf{v\sub{i}}-\itbf{v\sub{j}}$, and get
\begin{eqnarray}
	\itbf{v\sub{i}\super{\prime}}
	&=& \itbf{v\sub{i}}
	    - (\itbf{\hat{\sigma}}\cdot\itbf{v\sub{ij}})\itbf{\hat{\sigma}},
\nonumber\\
	\itbf{v\sub{j}\super{\prime}}
	&=& \itbf{v\sub{j}}
	    + (\itbf{\hat{\sigma}}\cdot\itbf{v\sub{ij}})\itbf{\hat{\sigma}},
\label{eq4}
\end{eqnarray}
with $\itbf{\hat{\sigma}}$ the unit vector in the direction of the
line connecting the center of the two disks at contact, i.e.,
$\itbf{\hat{\sigma}}=a^{-1}\itbf{r\sub{ij}}$.  Primed quantities
denote post-collisional values throughout this paper.  The positions
remain unchanged, $\itbf{r\sub{i}\super{\prime}}=\itbf{r\sub{i}}$,
$\itbf{r\sub{j}\super{\prime}}=\itbf{r\sub{j}}$.

In between collisions, the coordinates of disk~$l$ satisfy
\begin{equation}
	\dot{\itbf{r\sub{l}}} = \itbf{v\sub{l}}
\quad;\quad
	\dot{\itbf{v\sub{l}}}
	= \frac{1}{m} \itbf{F\sub{l}}(\{\itbf{r\sub{h}},\itbf{v\sub{h}}\}),
\label{eq5}
\end{equation}
The forces $\itbf{F\sub{l}}$ are smooth functions of the coordinates
$\{\itbf{r\sub{h}},\itbf{v\sub{h}}\}$, $h=1\ldots N$. In an
instantaneous collision between disks $i$ and $j$, the smooth forces
$\itbf{F\sub{i}}$ and $\itbf{F\sub{j}}$ cannot perform any action,
therefore Eq.~(\ref{eq4}) describing the collision, holds for any
$\itbf{F\sub{l}}$.

The dynamics of the deviations $\itbf{\delta r\sub{l}}$ and
$\itbf{\delta v\sub{l}}$ in collisionless flight are given by the
linearized version of Eq.~(\ref{eq5}),
\begin{eqnarray}
	\dot{\itbf{\delta r\sub{l}}} &=& \itbf{\delta v\sub{l}}
\nonumber\\
	\dot{\itbf{\delta v\sub{l}}}
	&=& \frac{1}{m}\sum_{h=1}^N \sum_{a=1}^{2}
	\left(
	\frac{\partial\itbf{F\sub{l}}}{\partial r_{h,a}}\delta r_{h,a}
	+
	\frac{\partial\itbf{F\sub{l}}}{\partial v_{h,a}}\delta v_{h,a}
	\right),
\label{eq6}
\end{eqnarray}
where $\delta r_{h,a}$ and $\delta v_{h,a}$ denote the $a$-th
component of $\itbf{\delta r\sub{h}}$ and $\itbf{\delta v\sub{h}}$,
respectively.

To find the collision dynamics, we use a method developed both by
Gaspard and Dorfman~\cite{GD} and by Dellago, Posch and
Hoover\cite{Dellago}. Here, we work out the dynamics for the general
system of Eqs.~(\ref{eq4}) and (\ref{eq5}).  For the equilibrium case
this was done in Refs.~\cite{inszasz,Dellago}.  The reference
trajectory and the adjacent trajectory are infinitesimally close so we
can assume that they have the same collision sequence. The most subtle
ingredient in the derivation of the collision dynamics of deviations,
is the time difference $\delta t$ between the $(i,j)$ collision on the
two adjacent trajectories. We set the time of the $(i,j)$ collision
equal to zero for the reference trajectory, so that of the adjacent
trajectory equals $\delta t$.  We consider here the case that $\delta
t$ is positive, but one easily checks that the final results are
equally valid for negative $\delta t$.  We define
\[
	\begin{array}{ll}
		\itbf{\delta r\sub{l}}
		= \itbf{r\sub{l}\super{*}}(0^-) - \itbf{r\sub{l}}(0^-),
	&
		\itbf{\delta v\sub{l}}
		= \itbf{v\sub{l}\super{*}}(0^-) - \itbf{v\sub{l}}(0^-),
	\\
		\itbf{\delta r\sub{l}\super{\prime}}
		= \itbf{r\sub{l}\super{*}}(\delta t^+)
		  -\itbf{r\sub{l}}(\delta t^+),
	&
		\itbf{\delta v\sub{l}\super{\prime}}
		= \itbf{v\sub{l}\super{*}}(\delta t^+)
		  -\itbf{v\sub{l}}(\delta t^+),
	\end{array}
\]
where $l=i$ or $j$, the superscript $*$ denotes values on the adjacent
trajectory and $+$ and $-$ indicate after and before collision,
respectively.

The time shift $\delta t$ can be found from the requirement that at
the instant of collision, the two disks are a distance $a$ apart,
i.e., $|\itbf{r\sub{ij}}(0)|=a$ and $|\itbf{r\sub{ij}\super{*}}(\delta
t)|=a$.  Because the time difference $\delta t$ is infinitesimal, we
only have to express the $\itbf{r\sub{l}\super{*}}(\delta t)$ to
linear order in $\delta t$, yielding $\itbf{r\sub{l}\super{*}}(\delta
t) = \itbf{r\sub{l}}+\itbf{\delta r\sub{l}} + \itbf{v\sub{l}}\delta
t$.  Note that here (and in the rest of this section) unprimed
quantities without time specification are assumed to carry their value
at $t=0$ before the collision, e.g., $\itbf{v\sub{l}} =
\itbf{v\sub{l}}(0^-)$.  From the requirement
$|\itbf{r\sub{ij}\super{*}}(\delta t)|^2 - |\itbf{r\sub{ij}}|^2=0$, we
get
\begin{eqnarray*}
          2\itbf{r\sub{ij}}\cdot
	     (\itbf{\delta r\sub{ij}}+\itbf{v\sub{ij}}\delta t)
	&=& 0
\end{eqnarray*}
to linear order in the deviations, so
\begin{equation}
        \delta t = - \frac{\itbf{r\sub{ij}}\cdot\itbf{\delta r\sub{ij}}}
                          {\itbf{r\sub{ij}}\cdot\itbf{v\sub{ij}}}
		=
		   - \frac{\itbf{\hat{\sigma}}\cdot\itbf{\delta r\sub{ij}}}
		          {\itbf{\hat{\sigma}}\cdot\itbf{v\sub{ij}}}.
\label{eq7}
\end{equation}
The difference in collision normal $\itbf{\delta\hat{\sigma}} =
\itbf{\hat{\sigma}\super{*}}-\itbf{\hat{\sigma}}$ follows from
\begin{eqnarray*}
	\itbf{\delta\hat{\sigma}}
	&=& \frac{\itbf{r\sub{ij}\super{*}}(\delta t) - \itbf{r\sub{ij}}}{a}
	 =\frac{\delta \itbf{r\sub{ij}}+\itbf{v\sub{ij}}\delta t}{a}
	 =  \frac{(\itbf{\hat{\sigma}}\cdot\itbf{v\sub{ij}})\identity
		 -\itbf{v\sub{ij}}\itbf{\hat{\sigma}}
             }
	     {a(\itbf{\hat{\sigma}}\cdot\itbf{v\sub{ij}})}
             \itbf{\delta r\sub{ij}},
\end{eqnarray*}
where we used Eq.~(\ref{eq7}) in the last equality. $\identity$ is the
identity matrix and we use the conventions that non-dotted products of
two vectors are dyadic products, and a product of matrices always
implies matrix multiplication, as does the product of a matrix with a
vector.

Consider first the position deviations. For the reference trajectory
we have, for $l=i$ or $j$,
\[
	\itbf{r\sub{l}}(\delta t)
	= \itbf{r\sub{l}}+\itbf{v\sub{l}\super{\prime}} \delta t,
\]
because the trajectory is determined by the velocity
$\itbf{v\sub{l}}(0^+)$ after the collision at $t=0$. For the adjacent
trajectory we have
\[
	\itbf{r\sub{l}\super{*}}(\delta t) = \itbf{r\sub{l}\super{*}}
  	+ \itbf{v\sub{l}\super{*}}(0^-) \delta t,
\]
because this trajectory has velocity $\itbf{v\sub{l}\super{*}}(0^-)$,
before colliding at time $\delta t$. We write
$\itbf{\delta r\sub{l}}'
=\itbf{r\sub{l}\super{*}}+\itbf{v\sub{l}\super{*}}\delta t
-\itbf{r\sub{l}}-\itbf{v\sub{l}\super{\prime}} \delta t$ and insert
the expressions for $\itbf{v\sub{l}\super{\prime}}$ from
Eq.~(\ref{eq4}) and the one for $\delta t$ from Eq.~(\ref{eq7}),
to find
\begin{eqnarray}
        \itbf{\delta r\sub{i}\super{\prime}} & = & \itbf{\delta r\sub{i}} -
        (\itbf{\delta 
r\sub{ij}}\cdot\itbf{\hat{\sigma}})\itbf{\hat{\sigma}},
\nonumber\\
        \itbf{\delta r\sub{j}\super{\prime}} & = & \itbf{\delta r\sub{j}} +
        (\itbf{\delta r\sub{ij}}\cdot\itbf{\hat{\sigma}})\itbf{\hat{\sigma}}
\label{eq8}
\end{eqnarray}
(neglecting again expressions quadratic in deviations).  For the
velocity deviation vectors on the reference trajectory, one finds for
$l=i$ or $j$,
\[
        \itbf{v\sub{l}}(\delta t^+)
	= \itbf{v\sub{l}}(0^+)
	  +\frac{1}{m}\itbf{F\sub{l}}\left(
	      \{\itbf{r\sub{h}},\itbf{v\sub{h}\super{\prime}}\}
	   \right)\delta t,
\]
Of course, only the velocities of the colliding disks,
$\itbf{v\sub{i}}$ and $\itbf{v\sub{j}}$, have really changed. We
abbreviate $\itbf{F\sub{l}} \left(
\{\itbf{r\sub{h}},\itbf{v\sub{h}\super{\prime}}\} \right)$ by
$\itbf{F\sub{l}\super{\prime}}$ and $\itbf{F\sub{l}} \left(
\{\itbf{r\sub{h}},\itbf{v\sub{h}}\} \right)$ by $\itbf{F\sub{l}}$ from
now on. For the adjacent trajectory,
\[
        \itbf{v\sub{l}\super{*}}(\delta t^+)
	= \left[\itbf{v\sub{l}\super{*}}
	  +\frac{1}{m} \itbf{F\sub{l}}
	  \left(\{\itbf{r\sub{h}\super{*}},\itbf{v\sub{h}\super{*}}\}\right)\delta 
t
	  \right]^{\prime}
        =\left[
	  \itbf{v\sub{l}\super{*}} + \frac{1}{m}\itbf{F\sub{l}}\delta t
	 \right]^{\prime}
\]
where in the last equation, the difference between
$\itbf{F\sub{l}\super{*}}\delta t$ and $\itbf{F\sub{l}}\delta t$ has
been neglected (second order in the deviations).  The prime means that
the collision rule for velocities should be applied. For the adjacent
trajectory, this is a collision with collision normal
$\itbf{\hat{\sigma}\super{*}}$, so, for disk~$i$,
\begin{eqnarray*}
        \itbf{v\sub{i}\super{*}}(\delta t^+)
	&=& \left[\itbf{v\sub{i}\super{*}}(\delta t^-)
		  -\itbf{\hat{\sigma}\super{*}}
		   \left(
			\itbf{\hat{\sigma}\super{*}}\cdot
			\itbf{v\sub{ij}\super{*}}(\delta t^-)
		   \right)
	    \right]
\\
        &=& \itbf{v\sub{i}}(\delta t^+)
             +\itbf{\delta v\sub{i}}
             -\itbf{\hat{\sigma}}\left(\itbf{\hat{\sigma}}\cdot
		\itbf{\delta v\sub{ij}}\right)
\\&&
	-
	[ (\itbf{\hat{\sigma}}\cdot\itbf{v\sub{ij}})\identity
             +\itbf{\hat{\sigma}}\itbf{v\sub{ij}}
          ]\itbf{\delta\hat{\sigma}}
\\&&
	+ \frac{1}{m}(\itbf{F\sub{i}}-\itbf{F\sub{i}\super{\prime}})\delta t
         - \frac{1}{m}\left(\itbf{\hat{\sigma}}\cdot
                \{\itbf{F\sub{i}}- \itbf{F\sub{j}}\}
                \right)\itbf{\hat{\sigma}}\delta t .
\end{eqnarray*}
Inserting the expressions for $\itbf{\delta\hat{\sigma}}$ and $\delta
t$, we obtain
\begin{eqnarray*}
        \itbf{\delta v\sub{i}\super{\prime}}
	&=& \itbf{\delta v\sub{i}}
            -\itbf{\hat{\sigma}}(\itbf{\hat{\sigma}}\cdot
                  \itbf{\delta v\sub{ij}})
	    - (\Q -{\cal E}_1) \itbf{\delta r\sub{ij}}
\end{eqnarray*}
where
\begin{eqnarray}
        \Q &=&
	\frac{[(\itbf{\hat{\sigma}}\cdot\itbf{v\sub{ij}})\identity
                 +\itbf{\hat{\sigma}}\itbf{v\sub{ij}}]
		  [(\itbf{\hat{\sigma}}\cdot\itbf{v\sub{ij}})\identity
                 -\itbf{v\sub{ij}}\itbf{\hat{\sigma}}]
                }
		{a(\itbf{\hat{\sigma}}\cdot\itbf{v\sub{ij}})},
\label{eq9}
\\
	{\cal E}_1
	&=& \frac{[\itbf{\hat{\sigma}}\cdot(\itbf{F\sub{i}}-\itbf{F\sub{j}})]
                   \itbf{\hat{\sigma}}
                   +\itbf{F\sub{i}\super{\prime}}-\itbf{F\sub{i}}
	    }{m}         
\frac{\itbf{\hat{\sigma}}}{\itbf{\hat{\sigma}}\cdot\itbf{v\sub{ij}}}.
\nonumber
\end{eqnarray}
Similarly, for disk $j$, we find
\begin{eqnarray}
        \itbf{\delta v\sub{j}\super{\prime}} &=& \itbf{\delta v\sub{j}} +
                \itbf{\hat{\sigma}}(\itbf{\hat{\sigma}}\cdot
                        \itbf{\delta v\sub{ij}})
			+
                        (\Q-{\cal E}_1) \itbf{\delta r\sub{ij}}
\label{eq10}
\end{eqnarray}
Thus when the dynamics of a point in phase space is determined by
Eqs.~(\ref{eq5}) and (\ref{eq4}), the dynamics of the deviation
vectors is given by Eqs.~(\ref{eq6}) and (\ref{eq8}-\ref{eq10}).

\section{Sheared system}
\label{sec:shear}

In the sheared hard sphere gas, it is convenient to transform to the
peculiar velocities of the particles,
\begin{equation}
       \itbf{V\sub{l}}=\itbf{v\sub{l}}-\itbf{u}(\itbf{r\sub{l}}) =
	\itbf{v\sub{l}}-\gamma y_l \itbf{\hat x}.
\label{eq11}
\end{equation}
The equations of motion in collisionless flight ($\dot{\itbf{r\sub
l}}=\itbf{v\sub{l}}$, $\dot{\itbf{v\sub{l}}}=0$) are transformed to
\begin{eqnarray}
    \dot{\itbf{r\sub{l}}} &=& \itbf{V\sub{l}} + \gamma y_l
	\itbf{\hat x} ,
\label{eq12}\\
        \dot{\itbf{V\sub{l}}} &=& - \gamma V_{l,y} \itbf{\hat x}
\label{eq13}
\end{eqnarray}
These equations are called the SLLOD equations of motion\cite{Evans}.
The pseudo-force $-m \gamma V_{l,y}\itbf{\hat{x}}$ is called the shear
force. Due to Eq.~(\ref{eq11}), in a collision between disks~$i$
and~$j$, the peculiar velocities transform as
\begin{eqnarray}
   \itbf{V\sub{i}\super{\prime}} &=& \itbf{V\sub{i}} - (\itbf{\hat{\sigma}}
                    \cdot \itbf{V\sub{ij}}) \itbf{\hat{\sigma}}
	-a\gamma\itbf{\hat{\sigma}}(\itbf{\hat x}\cdot\itbf{\hat{\sigma}})
		(\itbf{\hat y}\cdot\itbf{\hat{\sigma}}),
\nonumber\\
       \itbf{V\sub{j}\super{\prime}} &=& \itbf{V\sub{j}} + 
(\itbf{\hat{\sigma}}
                      \cdot \itbf{V\sub{ij}}) \itbf{\hat{\sigma}}
  +a\gamma\itbf{\hat{\sigma}}(\itbf{\hat x}\cdot\itbf{\hat{\sigma}})
		(\itbf{\hat y}\cdot\itbf{\hat{\sigma}}),
\label{eq14}
\end{eqnarray}
In simulations, boundary effects can be minimized by a special kind of
periodic boundary conditions, Lees-Edwards boundary conditions, in
which case the periodic copies in the $y$-direction are moving with
respect to another. When a particle crosses the boundary, it is put
back at the other end but with a corrected position and unchanged
{\em peculiar} velocity\cite{Cohentransport,Evans,LeesEdwards}.

The system as defined above does not have a steady state. The reason
is that the shear force continuously converts macroscopic kinetic
energy of flow into heat, i.e.\ internal kinetic energy.  In realistic
Couette-flow situations the work required for this is performed by the
shearing walls of the system. In the present situation this work is a
consequence of the
boundary conditions. Because the location of the y-coordinate of the
Lees-Edwards boundary is arbitrary, the system develops no temperature
gradient, in contrast to a system with realistic boundaries.  In such
a realistic system a stationary state is usually reached by the
establishment of a stationary heat flow towards the boundaries, which
absorb the heat and transmit it to a thermostat.  If one doesn't want
to include the environment explicitly, as is usually the case when one
performs MD-simulations, one needs to put in a mechanism by hand to
extract heat from the system.  Such a mechanism commonly is also
called a {\em thermostat}.  Several thermostats are around for
non-equilibrium systems, but we focus on one in particular. An extra
term is added to the equations of motion for the velocities during
collisionless flight, which become
\begin{equation}
        \dot{\itbf{V\sub{i}}} = - \gamma V_{yi} \itbf{\hat{x}}
        -\alpha\itbf{V\sub{i}}.
\label{eq15}
\end{equation}
We choose for $\alpha$ a constant positive value, so that the extra
terms can be interpreted as standard friction forces.  In molecular
dynamics simulations it is more common to use an isokinetic Gaussian
thermostat,\cite{Evans} which keeps the (peculiar) kinetic energy
$\sum_i\frac12m|\itbf{V\sub{i}}|^2$ strictly constant. In that case,
$\alpha$ depends on the positions and velocities of all the particles
and can be chosen such that the equations of motion are
time-reversible, i.e., form invariant under a time reversal
operation. $\alpha$ then may take both positive and negative values
and only on average will it be positive in the stationary state. In
the thermodynamical limit, this thermostat is equivalent to the one
with a constant $\alpha$,\cite{EvansSarman,mythermostat}\ and we
choose the latter one, as it is simpler. The value of $\alpha$
determines the average peculiar kinetic energy, which is identified
with the steady state temperature through
\begin{equation}
  \langle\sum_i\frac12m|\itbf{V\sub{i}}|^2\rangle = Nk_BT.
\end{equation}
The brackets here denote a time average, which, when the system is
ergodic, is also the average over an appropriate steady state
distribution. Our thermostat suppresses turbulence\cite{Evans} in
regimes where it is expected physically. So the model is
representative for an actual physical system only for low enough
Reynolds number.

In the low density regime we can use the Boltzmann equation for the
one-particle distribution function\cite{dorfmanbook,Cercignani}. We
only consider stationary flows of uniform shear, so we can focus on
the velocity distribution function of the form $f(\itbf{V})$, which is
normalized to unity. The Boltzmann equation for this distribution is
\begin{eqnarray}
&&  -\partial_{\itbf{V\sub{1}}}\cdot
\left[(\gamma V_{1y}\itbf{\hat x}+\alpha\itbf{V\sub 1})f_1\right]
\nonumber\\
&&
=
		\int\cdots\int' n a
|\itbf{\hat{\sigma}}\cdot\itbf{v\sub{12}}|
(f_1'f_2' - f_1 f_2)
d\itbf{V\sub{2}}\,d\itbf{\hat{\sigma}},
\label{eq16}
\end{eqnarray}
with the short hand notation $f_i=f(\itbf{V\sub{i}})$ and
$f_i'=f(\itbf{V\sub{i}}')$.  The prime on the integration denotes the
condition $\itbf{\hat{\sigma}} \cdot \itbf{v\sub{12}}<0$.

We want to discuss briefly how one can solve this Boltzmann equation
for small shear rates. In equilibrium, $\gamma=\alpha=0$, and the
right-hand side vanishes if $f$ is a Maxwell velocity distribution. To
allow for treating the left-hand side as a perturbation, $\gamma$ has
to be small compared to $na \langle|\itbf{v\sub{12}}|\rangle$. Hence,
the small parameter proportional to the shear rate is
$\tilde{\gamma}=\gamma a/(\tilde{n} v_0)$.
A Chapman-Enskog expansion of Eq.~(\ref{eq16}) can now be made by
expanding in powers of this parameter. The expansion of $f$ has the
form
\begin{equation}
f(\itbf{V}) = \varphi(\itbf{V}) [ 1+ \tilde{\gamma} g_1(\itbf{V}) +
\tilde{\gamma}^2 g_2(\itbf{V})+\ldots],
\label{eq17}
\end{equation}
where $\varphi$ is the Maxwell distribution
\begin{equation}
	\varphi(\itbf{V}) =  \frac{m}{2\pi k_B T}
	e^{-m|\smallitbf V|^2/(2k_BT)}.
\label{eq18}
\end{equation}

In the rest of the paper, we use $\tilde{\gamma}$ as an independent
variable, separate from the density $\tilde{n}$.  In this spirit,
higher density corrections are terms which vanish (relatively) when
the density is lowered while $\tilde{\gamma}$ is kept fixed.  Thus we
neglect the last terms $\pm a\gamma(\itbf{\hat{\sigma}}\cdot\itbf{\hat
x}) (\itbf{\hat{\sigma}}\cdot\itbf{\hat y})$ in Eq.~(\ref{eq14}), as
they are ${\cal O}(\tilde{\gamma}\tilde{n}v_0)$, i.e., one order in
density higher than the other terms, which are of ${\cal
O}(v_0)$. Likewise, the restriction on the integral in
Eq.~(\ref{eq16}) is replaced by
$\itbf{V\sub{12}}\cdot\itbf{\hat{\sigma}}<0$ and
$\itbf{u}(a\itbf{\hat{\sigma}})$ is neglected in
$|\itbf{\hat{\sigma}}\cdot\itbf{v\sub{12}}|=
|\itbf{\hat{\sigma}}\cdot(\itbf{V\sub{12}}+\itbf{u}(a\itbf{\hat{\sigma}}))|$.
Finally, $\itbf{v\sub{ij}}$ is replaced by $\itbf{V\sub{ij}}$ in the
matrix $\Q$ in Eq.~(\ref{eq9}).

Within the framework of the Chapman-Enskog expansion the zeroth order
solution, $\varphi$, entirely determines the temperature $T$; higher
orders should not change the second moment of $f$.  For fixed $\gamma$
and $T$ the friction coefficient $\alpha$ has to be expanded as
$\alpha = \gamma [\alpha_{1} + \tilde{\gamma} \alpha_{2} +\ldots]$.
To determine the coefficients, we note that $\langle
\partial_t\sum_i|\itbf{V\sub{i}}|^2\rangle=0$ in the stationary state
and that $\sum_i|\itbf{V\sub{i}}|^2$ is conserved in a collision, so
that for the time derivative we can insert Eq.~(\ref{eq15}). In this
way, we see that
\begin{equation}
  \alpha = \gamma \langle V_xV_y\rangle/\langle|\itbf{V}|^2\rangle
  = \frac{\gamma m}{k_BT}\int V_xV_y f(\itbf{V}) \,d\itbf{V}.
\label{eq19}
\end{equation}
Thus, if we know $f$ up to order $\tilde{\gamma}^{n-1}$, we can
calculate $\alpha$ up to order $\tilde{\gamma}^n$, which we need to
calculate $f$ up to order $\tilde{\gamma}^n$. One also immediately
sees that $\alpha_1=0$, because the average of the odd function
$V_xV_y$ with the even distribution $\varphi$ is zero. This is no
surprise, as $\alpha$ gives the dissipation and this should be an even
function of $\gamma$.

We will need the distribution $s(\itbf{V},\tau;t)$ of particles at
time $t$ with velocity $\itbf{V}$ and time $\tau$ passed since their
last collision.  In the stationary state this satisfies the equation
\[
	\partial_{\itbf{V}}\cdot
	[\dot{\itbf{V}}\,
	s
	(\itbf{V},\tau)]
	+\partial_{\tau}
	s
	(\itbf{V},\tau)
	=-\nu(\itbf{V})
	s
	(\itbf{V},\tau),
\]
for $\tau>0$. The collision frequency $\nu$ is given by
\begin{equation}
	\nu(\itbf{V\sub{1}}) =
		\int\cdots\int' na
	|\itbf{\hat{\sigma}}\cdot\itbf{V\sub{12}}|
	f_2 d\itbf{V\sub{2}} d\itbf{\hat{\sigma}}.
	\label{eq20}
\end{equation}
We rewrite the density $s$ in terms of a conditional distribution
function:
$s(\itbf{V},tau) = f(\itbf{V}) S(\tau|\itbf{V})$. With
\begin{eqnarray}
	\nu^* (\itbf{V})
	 &\equiv&\nu(\itbf{V})+\partial_{\itbf{V}}\cdot\dot{\itbf{V}}
	 +\dot{\itbf{V}}\cdot\partial_{\itbf{V}}\ln f,
\label{eq21}
\end{eqnarray}
we find the following equation for $S(\tau|\itbf{V})$,
\[
	\partial_\tau
	S
	(\tau|\itbf{V}) +
	\dot{\itbf{V}}\cdot\partial_{\itbf{V}}
	S
	(\tau|\itbf{V})
	= -\nu^*(\itbf{V})
	S
	(\tau|\itbf{V}),
\]
with the initial condition to be determined by normalization.  The
general solution of this equation is
\begin{eqnarray*}
	S
	(\tau|\itbf{V}) &=&
	S_0(\itbf{V}(-\tau))
	\exp\left[
		- \int_{-\tau}^0  \nu^*(\itbf{V}(t')) dt'
	\right]
\end{eqnarray*}
where $\itbf{V}(t)$ is the solution of the equations of motion,
Eq.(\ref{eq15}), with initial condition $\itbf{V}(0)=\itbf{V}$, i.e.,
$\itbf{V}(t) = e^{-\alpha t}[\itbf{V} - \itbf{\hat{x}} \gamma t V_y]$.
It is simple to show that for $\int_0^\infty S(\tau|\itbf{V}) d\tau$
to be equal to $1$, we need $S_0 (\itbf{V}) = \nu^* (\itbf{V})$, so
\begin{eqnarray}
	S(\tau|\itbf{V}) &=&
	\nu^*(\itbf{V}(-\tau))
	\exp\left[
		- \int_{-\tau}^0  \nu^*(\itbf{V}(t')) dt'
	\right]
.
\label{eq22}
\end{eqnarray}

\section{Generalized Boltzmann Equation}

\label{sec:EBE}

In this section we derive a generalized Boltzmann equation for the
distribution function of $\itbf{V}$ and $\itbf{\delta V}$.  But first
of all, we make the dynamics of the deviations explicit for the case
of the SLLOD equations (\ref{eq12}--\ref{eq15}) that we are
investigating.  In terms of $\itbf{r}_i$ and $\itbf{v}_i$, we have
$\itbf{\dot{v}\sub{i}}=\alpha\gamma y_i \itbf{\hat{x}} -
\alpha\itbf{v\sub{i}}$.  For this case, we get
\begin{eqnarray*}
  {\cal E}_1
  &=& \alpha\gamma a
\frac{\itbf{\hat{\sigma}}\cdot\itbf{\hat{x}}\,\itbf{\hat{\sigma}}\cdot\itbf{\hat{y}}}  
{\itbf{\hat{\sigma}}\cdot\itbf{v\sub{ij}}}\itbf{\hat{\sigma}}\itbf{\hat{\sigma}},
\end{eqnarray*}
(see Eq.~\ref{eq9}).  The $\itbf{\delta V}$ then change in a collision
according to
\begin{eqnarray}
        \itbf{\delta V\sub{i}\super{\prime}} &=& \itbf{\delta V\sub{i}}
                -\itbf{\hat{\sigma}}(\itbf{\hat{\sigma}}\cdot
                        \itbf{\delta V\sub{ij}})
                        -(\Q-{\cal E}_1-{\cal E}_2) \itbf{\delta r\sub{ij}}
\nonumber
\\
        \itbf{\delta V\sub{j}\super{\prime}} &=& \itbf{\delta V\sub{j}} +
                \itbf{\hat{\sigma}}(\itbf{\hat{\sigma}}\cdot
                        \itbf{\delta V\sub{ij}}) +
                        (\Q-{\cal E}_1-{\cal E}_2) \itbf{\delta
                r\sub{ij}}
\label{eq23}
\end{eqnarray}
where ${\cal E}_2=\gamma[\itbf{\hat
x}(\itbf{\hat{\sigma}}\cdot\itbf{\hat y})
\itbf{\hat{\sigma}}-\itbf{\hat{\sigma}}(\itbf{\hat
x}\cdot\itbf{\hat{\sigma}})\itbf{\hat y} ]$.  The terms ${\cal E}_1$
and ${\cal E}_2$ are of higher order in the density than $\Q$, which
is of order $(v_0/a)$, as one sees when expressing these quantities in
terms of $\tilde{\gamma}$ and $\tilde{n}$; ${\cal E}_1= {\cal
O}(v_0\tilde{\gamma}^3\tilde{n}^2/a)$ and ${\cal E}_2={\cal
O}(v_0\tilde{\gamma}\tilde{n}/a)$.  As we are restricting ourselves to
the leading powers in density, with at most correction terms of
relative order $1/|\ln \tilde{n}|$, we neglect ${\cal E}_1$ and ${\cal
E}_2$ in the sequel.

In collisionless flight, the dynamics of deviations is,
\begin{eqnarray*}
\begin{array}{rclrcl}
  \delta\dot{x}_i &=& \gamma\delta y_i +\delta V_{i,x};
&
   \delta\dot{y}_i& =& \delta V_{i,y};
\\
  \delta\dot{V}_{i,x} &=& -\gamma \delta V_{i,y} - \alpha
	\delta V_{i,x};
&
   \delta\dot{V}_{i,y} &=& - \alpha \delta V_{i,y}.
\end{array}
\end{eqnarray*}
The solutions of these equations are:
\begin{eqnarray}
   \itbf{\delta V\sub{i}}(t)
	&=&
	e^{-\alpha t}[ \itbf{\delta V\sub{i}}(0)
		- \itbf{\hat x} \gamma t \delta V_{y,i}(0) ]
\label{eq24}
\\
   \itbf{\delta r}_{i}(t)
	&=&
	\itbf{\delta r}_{i}(0) + \delta	y_i(0) \gamma t\itbf{\hat x}
		+\frac{1-e^{-\alpha t}}{\alpha} \itbf{\delta V\sub{i}}(0)
\nonumber\\
	&&
		+\frac{\alpha t(1+e^{-\alpha t})-2+2e^{-\alpha t}}
		{\alpha^2}
		\gamma \delta V_{i,y}(0)\itbf{\hat x}
\nonumber\\
	&\equiv&
		t {\bf S}_t \itbf{\delta V\sub{i}}(0)
		+
		\itbf{\delta r}_i(0)
		+\delta y_i\gamma t \itbf{\hat x}.
\label{eq25}
\end{eqnarray}

We follow the deviation dynamics from collision to collision.  So we
are interested in the case where the time $t$ in the equations above
is of the order of a mean intercollisional flight time, $t={\cal
O}(a/(v_0\tilde{n}))$. Then $\alpha t={\cal O} (\tilde{\gamma}^2)$,
and $\gamma t={\cal O}(\tilde{\gamma})$, and both are of zeroth order
in the density. We {\em assume} that
\begin{equation}
	{\cal O}(\itbf{\delta V\sub{i}})
	={\cal O}(\itbf{\delta r\sub{i}})\times v_0/a,
\label{eq26}
\end{equation}
just after a collision.  Then, at the next collision, the terms with a
factor $t\itbf{\delta V}_i (0)$ in Eq.~(\ref{eq25}) typically are one
order in $1/\tilde{n}$ larger than the corresponding terms
proportional to $\itbf{\delta r}_{i}(0)$, which therefore may be
neglected.  The $\itbf{\delta r}_i$ deviations just before a collision
then are $\itbf{\delta r}_i (\tau_i)=\tau_i{\bf
S}_{\tau_i}\itbf{\delta V\sub{i}}(0)$, where $\tau_i$ is the time from
the previous collision of particle $i$.  The effective equation
linking the $\itbf{\delta V\sub{i}}$ just after collision to their
values just after the previous collisions of the two particles, is
found from Eqs.~(\ref{eq23}) and (\ref{eq25}):
\begin{equation}
  \itbf{\delta V\sub{i}\super{\prime}}
        \approx-\itbf{\delta V\sub{j}\super{\prime}}
        \approx
                -\Q
                 (
                    \tau_i {\bf S}_{\tau_i} \itbf{\delta V\sub{i}}
                   -\tau_j {\bf S}_{\tau_j} \itbf{\delta V\sub{j}}
                 ),
\label{eq27}
\end{equation}
where we neglected terms of higher order in $\tilde{n}$ coming from
the center-of-mass contribution. To check for the consistency of our
assumption about the relative order of velocity and position
deviations just after a collision, we consider also $\itbf{\delta
r}_i$ and $\itbf{\delta r}_j$ after the next collision:
\[
\itbf{\delta r\sub{i}\super{\prime}}
        \approx
                -\itbf{\delta r\sub{j}\super{\prime}}
        \approx
		(\sfrac12\identity-\itbf{\hat{\sigma}}\itbf{\hat{\sigma}})
                 (
                    \tau_i {\bf S}_{\tau_i}\itbf{\delta V\sub{i}}
                   -\tau_j {\bf S}_{\tau_j}\itbf{\delta V\sub{j}}
                 ),
\]
and we see that if Eq.~(\ref{eq26}) holds for $\itbf{\delta
r\sub{i}}$, $\itbf{\delta V\sub{i}}$ and $\itbf{\delta r}_j$,
$\itbf{\delta V\sub{j}}$ just after the previous collision, it also
holds for $\itbf{\delta r\super{\prime}\sub{i}}$, $\itbf{\delta
V\sub{i}\super{\prime}}$ and $\itbf{\delta r\sub{j}\super{\prime}}$,
$\itbf{\delta V\sub j\super{\prime}}$.

In two dimensions, the matrix $\Q$ in Eq.~(\ref{eq27}) is
\[
\Q = \frac{({\bf R} \itbf{v\sub{ij}\super{\prime}})({\bf R} 
\itbf{v\sub{ij}})}
{a(\itbf{\hat{\sigma}}\cdot\itbf{v\sub{ij}})}
\]
where ${\bf R}$ denotes a rotation over 90 degrees counterclockwise.
To leading order in the density we may replace $\itbf{v\sub{ij}}$ by
$\itbf{V\sub{ij}}$. Neglecting further ${\cal E}_1$ and ${\cal E}_2$
in Eq.~(\ref{eq23}) we obtain as effective equations of change for the
velocity deviation vectors in a collision,
\begin{eqnarray}
        \itbf{\delta V}'_i & = & \frac{{\bf R} \itbf{V}'_{ij}
                        \left[{\bf R}
                        \itbf{V\sub{ij}}\cdot\left(
			\tau_j{\bf S}_{\tau_j}\itbf{\delta V\sub{j}}-
                        \tau_i{\bf S}_{\tau_i}\itbf{\delta V\sub{i}}\right)
			\right]
  }{a(\itbf{\hat{\sigma}}\cdot\itbf{V\sub{ij}})},
\label{eq28}
\\
        \itbf{\delta V}_j' & = & -\itbf{\delta V}_i'.
\nonumber
\end{eqnarray}
To compare different contributions to $\itbf{\delta V}'_i$ and
$\itbf{\delta V}'_j$, we want to know the order of $\itbf{\delta
V\sub{i}}$ and $\itbf{\delta V\sub{j}}$. We write the value of
$\itbf{\delta V\sub{i}}$ {\em just after the previous collision} as
\begin{equation}
        \itbf{\delta V\sub{i}}
              \equiv
                 v_0 
\left(\frac{1}{\tilde{n}}\right)^{k_i}\!\!\itbf{\hat{e}}_i
                        = v_0 e^{k_i/\vartheta} \itbf{\hat{e}}_i ,
\label{eq29}
\end{equation}
where $\itbf{\hat{e}}_i$ is a unit vector and $\vartheta$ is defined as
\[
	\vartheta \equiv \frac{1}{|\ln\tilde{n}|}
\]
The clock values and unit vectors thus defined do not change during
collisionless flight, only in collisions.  In contrast to earlier work
in Refs.~\cite{myself,inszasz}, the clock values $k_i$ are real
numbers here, not integers.

We obtain $k_1'$ from Eqs.~(\ref{eq28}) and (\ref{eq29}):
\[
  k_1' =\vartheta \ln\left|
\frac{
{\bf R}\itbf{V\sub{12}}\cdot{\bf 
S}_{\tau_2}\itbf{\hat{e}}_2\tau_2e^{k_2/\vartheta}
-{\bf R}\itbf{V\sub{12}}\cdot{\bf 
S}_{\tau_1}\itbf{\hat{e}}_1\tau_1e^{k_1/\vartheta}
}{a|\itbf{\hat{\sigma}}\cdot\itbf{\hat V}_{21}|}
\right|
\]
and distinguish two cases, telling us which of the two terms inside
the logarithm is the largest, and hence, the most important.  We
define
\begin{eqnarray}
   b(\itbf{V},\tau,\itbf{\hat{e}},\itbf{\hat{\sigma}})
	&=& \ln\left|
	\frac{\tau\tilde{n}{\bf R}\itbf{V}\cdot{\bf S}_{\tau}\itbf{\hat{e}}}
	     {a\itbf{\hat V}\cdot\itbf{\hat{\sigma}}}
	\right|,
\label{eq30}
\end{eqnarray}
and say that
\begin{itemize}
\item[$I$:] disk $1$ dominates disk $2$ if
\end{itemize}
\begin{equation}
k_1+\vartheta 
b(\itbf{V\sub{12}},\tau_1,\itbf{\hat{e}}_1,\itbf{\hat{\sigma}})
>k_2+\vartheta 
>b(\itbf{V\sub{12}},\tau_2,\itbf{\hat{e}}_2,\itbf{\hat{\sigma}})
\label{eq31}
\end{equation}
and that
\begin{itemize}
\item[$I\!I$:]
disk $2$ dominates disk $1$ otherwise.
\end{itemize}
We also define an ``alignment'' criterion:
\begin{itemize}
\item[($+$)] The unit vectors $\itbf{\hat{e}}_1$ and
$\itbf{\hat{e}}_2$ are said to be aligned if ${\bf
R}\itbf{V\sub{12}}\cdot{\bf S} _{\tau_1}\itbf{\hat{e}}_1$ has the same
sign as ${\bf R}\itbf{V\sub{12}}\cdot{\bf S}
_{\tau_2}\itbf{\hat{e}}_2$,
\item[($-$)]  and anti-aligned if they have opposite signs.
\end{itemize}
{}From now on, when a $\pm$ or $\mp$ occurs in an equation, the upper
sign corresponds to the aligned case, the lower one to the
anti-aligned case.  With these definitions, we write for the case $I$:
\begin{eqnarray}
	k_1' &=& k_1 +1 +\vartheta b(\itbf{V\sub{12}},
	\tau_1,\itbf{\hat{e}}_1,\itbf{\hat{\sigma}})
\nonumber\\&&		+
	\vartheta
\ln\left[1\mp e^{(k_2-k_1)/\vartheta}
\frac{\exp[b(\itbf{V\sub{12}},\tau_2,\itbf{\hat{e}}_2,\itbf{\hat{\sigma}})]}
{\exp[b(\itbf{V\sub{12}},\tau_1,\itbf{\hat{e}}_1,\itbf{\hat{\sigma}})]} 
\right]
\label{eq32}
\end{eqnarray}
and for case $I\!I$:
\begin{eqnarray}
k_1' &=& k_2 +1 +
\vartheta b(\itbf{V\sub{12}},\tau_2,\itbf{\hat{e}}_2,\itbf{\hat{\sigma}})
\nonumber\\&&
+ \vartheta \ln\left[1\mp e^{(k_1-k_2)/\vartheta}
\frac{\exp[b(\itbf{V\sub{12}},\tau_1,\itbf{\hat{e}}_1,\itbf{\hat{\sigma}})]}
{\exp[b(\itbf{V\sub{12}},\tau_2,\itbf{\hat{e}}_2,\itbf{\hat{\sigma}})]} 
\right]
\label{eq33}
\end{eqnarray}
At this point it can be made clear why the distinction of one disk
dominating over another was made. Consider Eq.~(\ref{eq32}). Because
$\vartheta$ is small for low densities, and because Eq.~(\ref{eq31})
holds, the term after ``$1\mp$'' inside the logarithm tends to be
small, at least if $k_1$ and $k_2$ differ by an amount of
$O(1)$. Consequently, this whole term is small, or, more precisely, it
is appreciable only for $|k_2-k_1|=O(\vartheta)$.  Also the term
$\vartheta b$ is typically small. Hence $k_1'=k_1+1$ almost always, in
the limit that $\vartheta\rightarrow0$ (the infinitely dilute gas).
The same limit in case $I\!I$ yields $k_1'=k_2+1$.  We see that
indeed, the ``dominant'' particle determines the value of the clocks
after collision, at least if the density is low enough.  This limiting
dynamics for low density, expressed in Eq.~(\ref{eq2}), was derived
before in Refs.~\cite{myself} and \cite{inszasz} and proved sufficient
for obtaining the leading term in the density expansion of the largest
Lyapunov exponent.

We consider the conditional probability distribution function for
having clock value $k$ and unit vector $\itbf{\hat{e}}$ just after a
collision at time $t$, given that the post-collisional velocity is
$\itbf{V}$. This function is denoted by
${\hat{f}}(k,\itbf{\hat{e}}|\itbf{V};t)$ and it obeys
\begin{equation}
  \omega (k,\itbf{\hat{e}},\itbf{V};t)
	=
		\omega(\itbf{V}) {\hat{f}}(k,\itbf{\hat{e}}|\itbf{V};t)
\label{eq34},
\end{equation}
where $\omega(X;t)$ stands for the rate at which particles with
attributes $X$ are produced in collisions at time t. The production
rate $\omega(\itbf{V})$ of $\itbf{V}$ in the stationary state is
independent of time and satisfies
\begin{equation}
\omega(\itbf{V\sub{1}})
	= \int\int' an|\itbf{\hat{\sigma}}\cdot\itbf{V\sub{12}}| f_1'f_2'
d \itbf{V\sub{2}} d \itbf{\hat{\sigma}} =
	\nu^*(\itbf{V\sub{1}})f_1,
\label{eq34a}
\end{equation}
where we used Eqs.~(\ref{eq15}), (\ref{eq16}), (\ref{eq20}) and
(\ref{eq21}).  By considering the rate of restituting collisions that
produce the right $(k,\itbf{\hat{e}},\itbf{V};t)$, we find that
$\omega(k$, $\itbf{\hat{e}}$, $\itbf{V};t)$ satisfies the equation
\begin{eqnarray}
\omega(k,\itbf{\hat{e}},\itbf{V};t)
	&=& \int\cdots\int'
		an|\itbf{\hat{\sigma}}\cdot\itbf{V\sub{12}}|
		 f_1f_2
			S(\tau_1|\itbf{V\sub{1}})
			S(\tau_2|\itbf{V\sub{2}})
\nonumber\\
	&&\times
	{\hat{f}}(k_1,\itbf{\hat{e}}_1|\itbf{V\sub{1}}(-\tau_1);t-\tau_1)
\nonumber\\&&\times
	{\hat{f}}(k_2,\itbf{\hat{e}}_2|\itbf{V\sub{2}}(-\tau_2);t-\tau_2)
\nonumber\\
	&&\times
		\delta(\itbf{V}_1'-\itbf{V})\delta(k_1'-k)
		\delta(\itbf{\hat{e}}_1'-\itbf{\hat{e}})
\nonumber\\
	&&\times
		 d \itbf{V\sub{1}} d \itbf{V\sub{2}} d k_1 d k_2
d \itbf{\hat{e}}_1 d \itbf{\hat{e}}_2 d \tau_1 d \tau_2 d 
\itbf{\hat{\sigma}}.
\label{eq35}
\end{eqnarray}
In the arguments of ${\hat{f}}$ the velocities need to be traced back
to the previous collision, because ${\hat{f}}$ was defined in terms of
the variables at that instant of time.  Hence the appearance of
$\itbf{V}(-\tau)$. The clock values and unit vectors do not need such
a correction, as they do not change in between collisions.  We
symmetrize the equation with respect to $\itbf{\hat{e}}$, because
$\itbf{\hat{e}}\rightarrow- \itbf{\hat{e}}$ only means interchanging
the reference and the adjacent trajectory, and this cannot affect
their rate of separation. Hence we can replace
$\delta(\itbf{\hat{e}}_1'-\itbf{\hat{e}})$ $=$ $\delta({\bf
R}\itbf{\hat V}_{12}'-\itbf{\hat{e}})$ by $[\delta({\bf R} \itbf{\hat
V}_{12}'-\itbf{\hat{e}})+\delta({\bf R}\itbf{\hat
V}_{12}'+\itbf{\hat{e}})]/2$ $=$
$\sfrac{1}{2}\delta(\itbf{\hat{e}}\cdot\itbf{\hat V}_{12}')$.

Through linear order in $\vartheta$ the logarithmic terms in the
expressions (\ref{eq32}) and (\ref{eq33}) for $k_1'$ may be ignored in
Eq.~(\ref{eq35}); their inclusion merely gives rise to corrections of
$O(\vartheta^2)$.  With this approximation Eq.~(\ref{eq35}) may be
rewritten as
\begin{eqnarray}
&& \omega(k,\itbf{\hat{e}},\itbf{V};t)
	\nonumber\\
&&=
		\frac{\partial}{\partial k}
		\int\cdots\int'
			S(\tau_1|\itbf{V\sub{1}})
			S(\tau_2|\itbf{V\sub{2}})
                        f_1f_2
\nonumber\\
	&&\quad\times
			an|\itbf{\hat{\sigma}}\cdot\itbf{V\sub{12}}|
                        \delta(\itbf{V}_1'-\itbf{V})
			\sfrac{1}{2}\delta(\itbf{\hat{e}}\cdot\itbf{\hat V}_{12}')
\nonumber\\
	&&\quad\times
                        C(k-1-\vartheta
	b(\itbf{V\sub{12}},\tau_1,\itbf{\hat{e}}_1,\itbf{\hat{\sigma}}),
			\itbf{\hat{e}}_1
			|\itbf{V\sub{1}}(-\tau_1);t-\tau_1)
\nonumber\\
	&&\quad\times
			C(k-1-\vartheta
	b(\itbf{V\sub{12}},\tau_2,\itbf{\hat{e}}_2,\itbf{\hat{\sigma}})
			,\itbf{\hat{e}}_2
			|\itbf{V\sub{2}}(-\tau_2);t-\tau_2)
\nonumber\\
	&&\quad\times
		 d \itbf{V\sub{1}} d \itbf{V\sub{2}} d \itbf{\hat{e}}_1 d \itbf{\hat{e}}_2
                 d \tau_1 d \tau_2 d \itbf{\hat{\sigma}}
		+ {\cal O}(\vartheta^2).
\label{eq36}
\end{eqnarray}
Here, we introduced a cumulative distribution,
defined as
\begin{equation}
C(k,\itbf{\hat{e}}|\itbf{V};t)=
\int_{-\infty}^{k}{\hat{f}}(k^*,\itbf{\hat{e}}|\itbf{V};t) d k^*.
\label{eq37}
\end{equation}
Integrating Eq.~(\ref{eq36}) from $-\infty$ to $k$, using
Eqs.~(\ref{eq34}), (\ref{eq34a}) and (\ref{eq37}) for the left hand
side, and changing integration variables from precollisional
velocities to post-collisional ones, we arrive at
\begin{eqnarray}
&&  \nu^*(\itbf{V\sub{1}}) C(k,\itbf{\hat{e}}|\itbf{V\sub{1}};t)
\nonumber\\&&
	=	\int\cdots\int'
			S(\tau_1|\itbf{V}_1')
			S(\tau_2|\itbf{V}_2')
                        f_1^{-1}f_1'f_2'
\nonumber\\
	&&\quad\times
                        C(k-1-\vartheta
		 b(\itbf{V}_{21}',\tau_1,\itbf{\hat{e}}_1,\itbf{\hat{\sigma}})
			,\itbf{\hat{e}}_1|\itbf{V}_1'(-\tau_1);t-\tau_1)
\nonumber\\
	&&\quad\times
                        C(k-1-\vartheta
	 b(\itbf{ V\sub{12}'} ,\tau_2,\itbf{\hat{e}}_2,\itbf{\hat{\sigma}})
			,\itbf{\hat{e}}_2|\itbf{V}_2'(-\tau_2);t-\tau_2)
\nonumber\\
	&&\quad\times
		an|\itbf{\hat{\sigma}}\cdot\itbf{V\sub{12}}|
		\sfrac{1}{2}\delta(\itbf{\hat{e}}\cdot\itbf{\hat V}_{12})
d \itbf{V\sub{2}} d \itbf{\hat{e}}_1 d \itbf{\hat{e}}_2 d \tau_1 d \tau_2 d
\itbf{\hat{\sigma}}.
\label{eq38}
\end{eqnarray}
This is the generalized Boltzmann equation for the cumulative
distribution function $C$, up to ${\cal O}(\vartheta^2)$, which will
be the starting point for our calculations of the maximal Lyapunov
exponent.

\section{Front propagation}
\label{sec:pulled}

The generalized Boltzmann equation (\ref{eq38}) can be interpreted as
describing a propagating front. The propagation here occurs on the
real line of clock values $k$: As clock values tend to grow there is a
movement towards higher clock values. The two "phases" that are
separated by the front are the stationary solutions $C(k,\itbf{\hat
e}|\itbf{V};t)\equiv0$ on the left (no particles have a clock value
smaller than the $k$-values in this region) and $C(k,\itbf{\hat
e}|;\itbf{V};t)=P(\itbf{\hat{e}}|\itbf{V})$ on the right (all
particles have a clock value smaller than the $k$-values in this
region).  $P(\itbf{\hat{e}}|\itbf{V})$ is the conditional probability
for a particle to have unit vector $\itbf{\hat{e}}$ given that its
velocity is $\itbf{V}$, i.e.
\begin{eqnarray*}
                    P(\itbf{\hat{e}}|\itbf{V\sub{1}})
        &&=    	\int\int' \frac{f_1'f_2'}
                {\nu^*(\itbf{V\sub{1}})f_1}
                        n a |\itbf{\hat{\sigma}}\cdot\itbf{V\sub{12}}|
                        \sfrac{1}{2}\delta(\itbf{\hat{e}}\cdot\itbf{\hat 
V}_{12})
                 d \itbf{V\sub{2}} d \itbf{\hat{\sigma}}.
\end{eqnarray*}
It is easy to see that $C\equiv 0$ is stable, whereas $C\equiv
P(\itbf{\hat{e}}|\itbf{V})$ is unstable. In the simplest situation,
the front has a fixed shape and moves to the right with a constant
velocity:
\begin{equation}
   C(k,\itbf{\hat{e}}|\itbf{V},t) = F(x,\itbf{\hat{e}}|\itbf{V})
\label{eq39}
\end{equation}
where $x = k - w\bar{\nu} t$ with $\bar{\nu}$ the average collision
frequency as mentioned already in the introduction.  The constant $w$
is called the {\em clock speed}. For the simpler clock model where
each particle is fully characterized by a single clock variable which
increases with time according to Eq.~(\ref{eq2}), it turned
out\cite{myself} that the front falls in the class of so-called pulled
fronts, as opposed to pushed fronts\cite{wimute}.  We will assume that
in the present model this is also the case.

\label{minimum}
In short, the asymptotic front speed of pulled fronts is determined as
follows. Insert a propagating front solution like Eq.~(\ref{eq39})
into the front equation~(\ref{eq38}), and linearize around the
unstable phase. The resulting linear equation has solutions which are
linear superpositions of exponential functions of the form $e^{-sx}$
(multiplied by a polynomial in $x$ in case of degeneracies). For fixed
$w$, there are a number of possible values of $s$ (typically infinite
but countable).  The dominant term in the superposition is the one
where $s=s_d$ has the smallest real part.  Since $C$ has to be
monotonic in $x$, the asymptotic large $x$ behavior $\sim e^{-s_d x}$
has to be so too, hence $s_d$ has to be real.  This turns out to be
possible only for $w$ larger than some critical value $w^*$. So the
asymptotic speed, if it exists, is greater than or equal to $w^*$. In
fact, for a large class of initial conditions, the asymptotic clock
speed is exactly $w^*$.  Especially, this is true for localized
initial conditions, implying that all initial $k$-values fall within a
finite range (or in fact, are smaller than some value $k_{max}$;
localization on the small-$k$ side is not important).  The same speed
is also selected by initial distributions that fall off sufficiently
rapidly (typically faster than any exponential of type $\exp[-c x]$)
at the large-$k$ side. For systems with finite numbers of particles,
which we are typically interested in, the initial distribution is
always localized, leading to automatic selection of the minimal clock
speed $w^*$. The additional effects of finite particle numbers only
reduce the clock speed further, because correlations between particles
tend to reduce their clock speed differences and thereby the boost a
"slow" particle receives from colliding with a "fast"
one\cite{inszasz}.  A more detailed exposition of the velocity
selection and other aspects of pulled and pushed fronts can be found
in Ref.~\cite{wimute}.

Applying this scheme to the generalized Boltzmann equation
Eq.~(\ref{eq38}), we first insert the Ansatz~(\ref{eq39}) and consider
the resulting equation.  We find
\begin{eqnarray*}
&&  \nu^*(\itbf{V\sub{1}}) F(x,\itbf{\hat{e}}|\itbf{V\sub{1}})
\nonumber\\&&	=\int\cdots\int'
			S(\tau_1|\itbf{V}_1')
			S(\tau_2|\itbf{V}_2')
                        f_1^{-1}f_1'f_2'
\nonumber\\
	&&\quad\times
  F(x-1-\vartheta b(\itbf{
	V\sub{12}}',\tau_1,\itbf{\hat{e}}_1,\itbf{\hat{\sigma}})
			+ w\bar{\nu}\tau_1,
			\itbf{\hat{e}}_1|\itbf{V}_1'(-\tau_1))
\nonumber\\
	&&\quad\times
  F(x-1-\vartheta 
b(\itbf{V\sub{12}}',\tau_2,\itbf{\hat{e}}_2,\itbf{\hat{\sigma}})
			+w\bar{\nu}\tau_2,
			\itbf{\hat{e}}_2|\itbf{V}_2'(-\tau_2))
\nonumber\\
	&&\quad\times
			an|\itbf{\hat{\sigma}}\cdot\itbf{V\sub{12}}|
			\sfrac{1}{2}\delta(\itbf{\hat{e}}\cdot\itbf{\hat V}_{12})
d \itbf{V\sub{2}} d \itbf{\hat{e}}_1 d \itbf{\hat{e}}_2
d \tau_1 d \tau_2 d \itbf{\hat{\sigma}}.
\end{eqnarray*}
We linearize this equation writing $F(x,\itbf{\hat{e}}|\itbf{V}) =
P(\itbf{\hat{e}}|\itbf{V\sub{1}})-
\Delta(x,\itbf{\hat{e}}|\itbf{V})$. The resulting linear equation for
$\Delta$ has superpositions of exponentials as solutions. It turns out
convenient to represent these as
\begin{eqnarray}
\Delta(x,\itbf{\hat{e}}|\itbf{V}) =
\sum a_i \frac{\nu^*(\itbf{V})+s_i
\bar{\nu}}{\nu^*(\itbf{V})}
\hat{A}_i(\itbf{\hat{e}},\itbf{V}) e^{-s_i x/w}.
\label{eq40}
\end{eqnarray}
The characteristic values $s_i$ and corresponding characteristic
functions $\hat{A}_i$ are solutions of the linearized equation with
$\Delta$ taking the form
\[
	\Delta(x,\itbf{\hat{e}}|\itbf{V}) =
	\frac{\nu^*(\itbf{V})+s\bar{\nu}}{\nu^*(\itbf{V})}
	\hat{A}(\itbf{\hat{e}},\itbf{V}) e^{-s x/w},
\]
This gives the following characteristic equation
\begin{eqnarray*}
&&
\Lambda(\nu^*(\itbf{V\sub{1}})+s\bar{\nu})\hat{A}(\itbf{\hat{e}},\itbf{V\sub{1}})
\nonumber\\
       &&=\int\cdots\int'
                        f^{-1}_1f_1'f_2'
                        an|\itbf{\hat{\sigma}}\cdot\itbf{V\sub{12}}|
                        \frac12\delta(\itbf{\hat{e}}\cdot\itbf{\hat 
V\sub{12}})
\nonumber\\
        &&\qquad\times
		\sum_{k=1,2}
                        P_s(\tau|\itbf{V}_k')
                        \exp\left[
                          -(\ln\Lambda)
	                \vartheta
			b(\itbf{V}_{21}',\tau,\itbf{\hat{e}}^*,\itbf{\hat{\sigma}})
			\right]
\nonumber\\
        &&\qquad\times
                        \hat{A}(\itbf{\hat{e}}^*,\itbf{V}_k'(-\tau))
                  d \itbf{V\sub{2}} d \itbf{\hat{e}}^* d \tau d 
\itbf{\hat{\sigma}}
,
\end{eqnarray*}
where we defined the eigenvalue $\Lambda=e^{-s/w}$ and
\begin{eqnarray}
   P_s(\tau|\itbf{V})
        &\equiv&(\nu^*(\itbf{V}(-\tau))+s\bar{\nu})
\nonumber\\
&&
\times
\exp\left[-\int_{-\tau}^{0}(\nu^*(\itbf{V}(t))+s\bar{\nu}
		)dt\right].
\label{eq41}
\end{eqnarray}
We make one more expansion in $\vartheta$:
\begin{equation}
  \Lambda (\nu^*+s\bar\nu) \hat{A}
        =
                \hat{\bf L}^{0}\hat{A}
                -\vartheta\ln\Lambda
                \hat{\bf L}^{1}\hat{A}
\label{eq42},
\end{equation}
where we left out the argument of $\nu^*$.
The operators $\hat{\bf L}^{0}$ and $\hat{\bf L}^{1}$ are
defined by
\begin{eqnarray}
&&  \left[\hat{\bf L}^{0}\hat{A}\right](\itbf{\hat{e}},\itbf{V\sub{1}})
		=
		\int\cdots\int'
                        f_1'f_2'f^{-1}_1
                      na|\itbf{\hat{\sigma}}\cdot\itbf{V\sub{12}}|
                        \frac12\delta(\itbf{\hat{e}}\cdot\itbf{\hat 
V}_{12})\nonumber\\
&&     \times
                 \sum_{k=1,2}
                        P_s(\tau|\itbf{V}_k')
		\hat{A}(\itbf{\hat{e}}^*,\itbf{V}_k'(-\tau))
		d\tau^* d \itbf{\hat{e}}^*
                 d \itbf{V\sub{2}} d \tau d \itbf{\hat{\sigma}}
,
\label{eq43}
\\
&&
  \left[\hat{\bf L}^{1}\hat{A}\right](\itbf{\hat{e}},\itbf{V\sub{1}})
  =\int\cdots \int'
                      f_1'f_2'f^{-1}_1
                      na |\itbf{\hat{\sigma}}\cdot\itbf{V\sub{12}}|
                        \frac12\delta(\itbf{\hat{e}}\cdot\itbf{\hat 
V}_{12})\nonumber\\
&&   \times
		   \sum_{k=1,2}
                        P_s(\tau|\itbf{V}_k')                         
b(\itbf{V}_{21}',\tau,\itbf{\hat{e}}^*,\itbf{\hat{\sigma}})
                        \hat{A}(\itbf{\hat{e}}^*,\itbf{V}_k'(-\tau)) d\tau^*
                 d \itbf{V\sub{2}} d \itbf{\hat{e}}^* d \tau d 
\itbf{\hat{\sigma}}
.
\label{eq44}
\end{eqnarray}
Finally, Eq.~(\ref{eq42}) can be transformed into an equation in
which only the
\label{page9}
integrated function over $\itbf{\hat{e}}$  enters, i.e.,
$A(\itbf{V}) \equiv [\bar {\bf{P}} \hat{A}](\itbf{V})
\equiv (2\pi)^{-1}\int d\itbf{\hat{e}}\, \hat{A}(\itbf{\hat{e}},\itbf{V})$.
Defining
\begin{eqnarray*}
	\left[{\bar{\bf L}}^{0} A\right](\itbf{\hat{e}},\itbf{V\sub{1}})
        &\equiv&\int\cdots\int'
                        f_1'f_2'f^{-1}_1
			na
                      |\itbf{\hat{\sigma}}\cdot\itbf{V\sub{12}}|
\pi\delta(\itbf{\hat{e}}\cdot\itbf{\hat V\sub{12}})
\nonumber
\\
        &&\times
                        \sum_{k=1,2}
                        P_s(\tau|\itbf{V}_k') A(\itbf{V}_k'(-\tau))
                 d \itbf{V\sub{2}} d \tau d \itbf{\hat{\sigma}}
,
\end{eqnarray*}
we see from Eq.~(\ref{eq43}) that we can write
\begin{equation}
   \hat{\bf L}^{0} =  \bar{\bf L}^{0} \bar{\bf P}.
\label{eq45}
\end{equation}
We see from Eqs.~(\ref{eq42}) and (\ref{eq45}) that $\hat{A} =
\Lambda^{-1}(\nu^* + s\bar\nu)^{-1}\bar{\bf L}^{0} A$ up to first
order in $\vartheta$, so Eq.~(\ref{eq42}) can be written as
\[
	\Lambda(\nu^* +s\bar\nu)\hat{A} = \bar{\bf L}^{0} A - \vartheta
		\frac{\ln\Lambda}{\Lambda} \hat {\bf L} (\nu^*+s\bar\nu)^{-1}
		\bar{\bf L}^0 A + O(\vartheta^2).
\]
Applying $\bar{\bf P}$ yields a closed
equation for $A$:
\begin{equation}
	\Lambda(\nu^* + s\bar\nu )A = {\bf L}^{0} A -
	\vartheta\Lambda^{-1}\ln\Lambda {\bf L}^{1} A,
\label{eq46}
\end{equation}
where
\[
	{\bf L}^{0} = \bar{\bf P}\bar{\bf L}^{0},
\]
and
\begin{equation}
	{\bf L}^{1} = \bar{\bf P}\hat{\bf L}^{1} (\nu^* + s\bar\nu)^{-1}
		      \bar{\bf L}^{0}
.
\label{eq47}
\end{equation}
These are just the first few operators that appear in an expansion in
$\vartheta$, each still containing all orders of the shear rate
$\tilde{\gamma}$.  We use Eq.~(\ref{eq46}) to find $\Lambda$ as a
function of $s$ through linear order in $\vartheta$.  As $\Lambda$
will be equated to $e^{-s/w}$, we are interested in the largest
eigenvalue $\Lambda(s)$ for which Eq.~(\ref{eq46}) can be satisfied,
because this corresponds to the most slowly decaying mode in
Eq.~(\ref{eq40}).  There are real solutions to $\Lambda(s)=e^{-s/w}$
for $s$ if $w>w^*$, for $w<w^*$ only complex solutions exist, leading
to undesirable oscillations. To find $w^*$ we look at the function
\begin{equation}
	\tilde w(s)
= - \frac s{\ln\Lambda(s)}
\label{eq48}
\end{equation}
There are no real solutions of $\tilde w(s)=w$ if $w$ is smaller than
the minimum of the function $\tilde w$, hence this minimum is $w^*$.

\section{Perturbation in the Density}
\label{sec:pertdens}

This section will be devoted to finding the solution of
Eq.~(\ref{eq46}) in a perturbation expansion in $\vartheta$, that is
we will find $\Lambda(s)$ and with that determine the minimum of
$-s/\ln\Lambda(s)$, which is the clock speed $w$.

We consider Eq.~(\ref{eq46}) for the largest eigenvalue $\Lambda(s)$.
Let us assume that Eq.~(\ref{eq46}) is solved to zeroth order by
$A^{0}$ and $\Lambda^{0}(s)$, i.e.
\begin{equation}
\Lambda^{0}(s) (\nu^* + s\bar\nu)A^{0} = {\bf L}^{0} A^{0}.
\label{eq48a}
\end{equation}
$A^{0}$ depends on $s$, although we will not denote this explicitly.  We
remark that this zeroth order equation was solved in
Ref.~\cite{inszasz} for the equilibrium case. In the general case,
${\bf L}^{0}$ is not self-adjoint, so we need the left
eigenfunction too. We denote
this function by $\bar A^{0}$.
An inner product is defined as
\begin{equation}
	(A,B) = \int A(\itbf{V}) B(\itbf{V}) \varphi(\itbf{V}) d\itbf{V} ,
\label{eq49}
\end{equation}
where the Maxwell distribution $\varphi$ was given in
Eq.~(\ref{eq18}).
The function $\bar A^{0}$ may be chosen such that under this inner product 
$(\bar A^{0},A^{0})=1$.
Inserting the expansions $A=A^{0} + \vartheta
A^{1} + O(\vartheta^2)$ and $\Lambda(s) = \Lambda^{0}(s) +
\vartheta\Lambda^{1}(s) + O(\vartheta^2)$ into Eq.~(\ref{eq46})
and considering the $O(\vartheta)$ terms, we get
\begin{eqnarray*}
&&  \Lambda^{0}(s)(\nu^* + s\bar\nu)A^{1} +
	\Lambda^{1}(s)(\nu^* +s\bar\nu)A^{0}
\\&&
	= {\bf L}^0(s) A^{1} -
	\frac{\ln\Lambda^{0}(s)}{\Lambda^{0}(s)} {\bf L}^{1} A^{0}.
\end{eqnarray*}
The inner product of this equation with $\bar A^{0}$ yields
\begin{equation}
	\Lambda^{1}(s) = -\frac{\ln\Lambda^{0}(s)}{\Lambda^{0}(s)}
	      \frac{
			\big(\bar A^{0} ,{\bf L}^{1}  A^{0}\big)
			}{
			\big(\bar A^{0} ,(\nu^* +s\bar\nu)
			A^{0}\big)
			},
\label{eq50}
\end{equation}
where we used Eq.~\ref{eq48a}.
The critical value $w$ is found from $(d\tilde w/ds)(s^*) = 0$ and $w=
\tilde w(s^*)$ (where $s^*$ is the location of the minimum).  Using
Eq.~(\ref{eq50}) for $\Lambda^{1}(s)$ and Eq.~(\ref{eq48})
we find $\tilde w(s) = \tilde w^{0}(s) +\vartheta \tilde
w^{1}(s)+{\cal O}(\vartheta^2)$, where
\begin{eqnarray}
	\tilde w^{0}(s) &=& -\frac{s}{\ln\Lambda^{0}(s)},
\nonumber
\\
	\tilde w^{1}(s) &=&
\frac{\tilde w^{0}(s)}{{\Lambda^{0}(s)}^2}
		\frac{
		\big( \bar A^{0}, {\bf L}^{1} A^{0}\big)
		}{
		\big( \bar A^{0}, (\nu^* + s\bar\nu) A^{0}\big)
		}.
\label{eq51}
\end{eqnarray}
Note that $\bar A^{0}$, $A^{0}$ and ${\bf L}^{1}$ depend on $s$ as well.
Again we assume the minimum of $\tilde w^{0}$ to be known, and to be
located at $s^{0}$, with a value of $w^{0}=\tilde w^{0}(s^{0})$. The
first order value of the minimum, which to first order is located at
$s^*=s^{0}+\vartheta s^{1}$, is (up to that order)
\[
\tilde w^{0}\big(s^{0}\big) + \frac{d\tilde w^{0}}{d s^{0}} \vartheta s^{1}
		+ \vartheta \tilde w^{1}\big(s^{0}\big)
	= \tilde w^{0}(s^{0}) + \vartheta \tilde w^{1}(s^{0}),
\]
where we used that the derivative of $\tilde w^{0}$ at $s^{0}$ is zero.
{}From Eq.~(\ref{eq51}) and
the identity $\Lambda^{0}=\exp(-s^{0}/w^{0})$,
the value of the critical clock speed up to order $\vartheta$
follows as
\[
	w
	= w^{0}
	\left[ 1 + \vartheta
		e^{2s^{0}/w^{0}}
		\frac{
		\big( \bar A^{0}, {\bf L}^{1} A^{0}\big)
		}{
		\big( \bar A^{0}, (\nu^* + s^{0}\bar\nu) A^{0}\big)
		}
		+ O(\vartheta^2)
	\right].
\]
This is the value of the clock speed that enters
into the Lyapunov exponent, according to Eq.~(\ref{eq3}), so
\begin{equation}
	\lambda^+ = w^{0}\bar{\nu} \left[
			\ln \frac1{\tilde{n}}
		+
		e^{2s^{0}/w^{0}}
		\frac{
		\big( \bar A^{0}, {\bf L}^{1} A^{0}\big)
		}{
		\big( \bar A^{0}, (\nu^* + s^{0}\bar\nu) A^{0}\big)
		}
		\right].
\label{eq52}
\end{equation}
Correction terms to this expression are $O(\bar\nu\vartheta)$.

\section{Equilibrium case}
\label{sec:equilibrium}

We will now explicitly evaluate Eq.~(\ref{eq52}) for the equilibrium
case.  In that case, i.e., $\gamma=0$, Eqs.~(\ref{eq43}) and
(\ref{eq44}) simplify strongly. One simplification is that the
velocity distribution is known to be the Maxwellian, i.e.,
$f(\itbf{V})=\varphi(\itbf{V})$, so in the integrand,
$f^{-1}_1f_1'f_2'=\varphi_2$.  Furthermore $\dot{\itbf{V}}=0$, which,
according to Eq.~(\ref{eq21}), makes $\nu^*(\itbf{V})$ and
$\nu(\itbf{V})$ equal to the equilibrium collision frequency
$\nu_0(\itbf{V})$ of a particle with velocity $\itbf{V}$. The
expression in Eq.~(\ref{eq41}) now gives
\[
   P_s(\tau|\itbf{V})
       =
                (\nu_0(\itbf{V})+s\bar{\nu})
                \exp\left\{-[\nu_0(\itbf{V})+s\bar{\nu}_0]\tau\right\}
.
\]
The expression for the function $b$ in Eq.~(\ref{eq30})
simplifies because ${\bf S}_t=\identity$:
\[
b(\itbf{V},\tau,\itbf{\hat{e}},\itbf{\hat{\sigma}}) =
\ln\left(\frac{\tau\tilde{n}|{\bf R}\itbf{V}\cdot\itbf{\hat{e}}|}
	{a|\itbf{\hat V}\cdot\itbf{\hat{\sigma}}|}\right).
\]
We see that now the integration over $\tau$ in Eqs.~(\ref{eq43}) and
(\ref{eq44}) can be performed.  The result is that Eq.~(\ref{eq46})
becomes
\begin{equation}
\Lambda_{0} (\nu_0+s\bar\nu_{0}) A_{0} = {\bf L}^{0}_{0} A_{0}
  -\vartheta\frac{\ln \Lambda_{0}}{\Lambda_{0}}{\bf L}^{1}_{0} A_{0}
\label{eq53}
\end{equation}
where
\[
   \left[{\bf L}^{0}_{0} A \right](\itbf{V\sub{1}})
	=\int\cdots\int'
                na
                |\itbf{\hat{\sigma}}\cdot\itbf{V\sub{12}}|
                 \left[A_1' + A_2'\right]
		 \varphi_2 d \itbf{V\sub{2}} d \itbf{\hat{\sigma}},
\]
Here $A'_k=A(\itbf{V}_k')$,
and, analogously to Eq.~(\ref{eq47}),
\[
{\bf L}^{1}_{0} = \bar{\bf P}\hat{\bf L}^{1}_{0}
	(\nu_{0} + s\bar\nu_{0})^{-1}
      \bar{\bf L}^{0}_{0},
\]
where
\begin{eqnarray*}
\left[\bar{\bf L}^{0}_{0} A \right](\itbf{\hat{e}},\itbf{V\sub{1}})
	&=&\int\cdots\int'
                     na|\itbf{\hat{\sigma}}\cdot\itbf{V\sub{12}}|
                        \sum_{k=1,2}  \pi A_k'
\nonumber
\\
        &&\times
                        \delta(\itbf{\hat{e}}\cdot\itbf{\hat V}_{12})
                        \varphi_2
d \itbf{V\sub{2}} d \itbf{\hat{\sigma}}
\end{eqnarray*}
\begin{eqnarray*}
\left[\hat{\bf L}^{1}_{0} \hat{A} \right](\itbf{\hat{e}},\itbf{V\sub{1}})
&=&\int\cdots\int'
\frac12 na{|\itbf{\hat{\sigma}}\cdot\itbf{V\sub{12}}|}
\sum_{k=1,2}
\hat A(\itbf{\hat{e}}^*,\itbf{V}_k')
\nonumber\\&&
\times
\ln\left[\frac{nae^{-\euler}|\itbf{V\sub{12}}|}{\nu_0(\itbf{V}_k')+s\bar{\nu}_0}
	\frac{|{\bf R}\itbf{\hat V}_{12}\cdot\itbf{\hat{e}}^*|}
		{|\itbf{\hat V}_{12}\cdot\itbf{\hat{\sigma}}|}\right]
\nonumber\\&&
\times
     \delta(\itbf{\hat{e}}\cdot\itbf{\hat V}_{12})
\varphi_2
d \itbf{V\sub{2}} d \itbf{\hat{\sigma}} d \itbf{\hat{e}}^*
\end{eqnarray*}
Here $\euler$ is Euler's number, $0.577\ldots$.  In general,
subscripts $0$ denote the equilibrium values of quantities introduced
before.

To zeroth order in $\vartheta$, Eq.~(\ref{eq53}) reads
\[
	\Lambda^{0}_{0}(\nu_{0} +s\bar\nu_{0})A^{0}_{0}
	= {\bf L}^{0}_{0} A^{0}_{0}.
\]
This is the very same eigenvalue problem that was found in a more
heuristic derivation of a Boltzmann equation for clock values in
Refs.~\cite{inszasz}, with exactly the same operators, except that $s$
was represented as $\gamma w$ and ${\bf L}^{0}_{0}$ and
$\nu_{0}+s\bar\nu_{0}$ were denoted as $\mbox{\boldmath$L$}$ and
$\bar\nu_{0}\mbox{\boldmath$W$}_s$, respectively.  The largest
eigenvalue $\Lambda^{0}_{0}(s)$ was determined numerically in that
paper, for a range of $s$, such that the minimum of $\tilde w$ could
be determined.  Briefly, the method used was the following.  A basis
of functions was constructed that are orthogonal with respect to the
inner product defined in Eq.~(\ref{eq49}), starting with $1$,
$\itbf{V}/v_0$, $\sfrac12|\itbf{V}/v_0|^2-1$ and
$\sfrac18|\itbf{V}/v_0|^4-|\itbf{V}/v_0|^2+1$. Finite matrices were
constructed containing the matrix elements of the operators with
respect to the four mentioned basis-functions. These finite matrices
were used in Eq.~(\ref{eq53}) instead of the real operators, to get a
numerically feasible eigenvalue problem.  In Ref.~\cite{inszasz} it
was checked that omitting the basis function containing $|\itbf{V}|^4$,
only changes the result by a few tenth of percents, so that is the
accuracy of the numerical results.  In this paper we choose to work
with just $1$, $\itbf{V}/v_0$, and $\sfrac12|\itbf{V}/v_0|^2-1$, as this
simplifies the calculations.  On that truncated basis the value and
position of the minimum are found to be
\begin{eqnarray}
	w^{0}_{0}
	&\approx& 4.732,
\label{eq54}
\\
	s^{0}_{0} &\approx& 3.506.
\label{eq55}
\end{eqnarray}
and the eigenfunction $A^{0}_{0}$ is
\begin{equation}
	A^{0}_{0}(\itbf{V}) \approx 0.612+0.194v_0^{-2}|\itbf{V}|^2 .
\label{eq56}
\end{equation}
The value of $w^{0}_{0}$ gives the
leading behavior of the largest Lyapunov exponent:
$\lambda^+_0 = - w^{0}_{0} \bar \nu_0  \ln \tilde{n} + O(1)$.

Using these results we can obtain the first correction term. For that, we 
adapt
Eq.~(\ref{eq52}) to this case:
\[
	\lambda^+ = w^{0}_{0}\bar{\nu}_{0} \left[
			\ln \frac1{\tilde{n}}
		\!+\!
		\exp{\bigg(\frac{2s^{0}_{0}}{w^{0}_{0}}\bigg)}
		\frac{
		( A^{0}_{0}, {\bf L}^{1}_{0} A^{0}_{0})
		}{
	( A^{0}_{0}, (\nu_{0} + s^{0}_{0}\bar\nu_{0}) A^{0}_{0})
		}
		\right].
\]
Note that ${\bf L}^{0}_{0}$ is self adjoint and therefore the left and
right eigenvectors are the same.  We have calculated the matrix
elements $(A^{0}_{0}, {\bf L}^{1}_{0} A^{0}_{0})$ and $(A^{0}_{0},
(\nu_{0} + s^{0}_{0}\bar\nu_{0}) A^{0}_{0})$ numerically respectively
analytically, using the numbers of Eqs.~(\ref{eq55}), (\ref{eq56}) and
(\ref{eq54}).  The numerical integration uses a sampling from the
Maxwellian and subsequent averaging of the rest of the integrand
appearing in the matrix elements (including the collision frequency,
for which we used a numerical approximation).  The results are $ (
A^{0}_{0}, {\bf L}^{1}_{0} A^{0}_{0})= -10.85 \,nav_0$ and $(
A^{0}_{0}, (\nu_{0} + s^{0}_{0}\bar\nu_{0}) A^{0}_{0})= 19.28 \,nav_0$
Thus we obtain the result that
\begin{equation}
  \lambda^+_0 = 4.732 \bar{\nu}_{0}[-\ln\tilde{n} -2.48
	+ O(1/\ln\tilde n)] .
\label{eq57}
\end{equation}

\section{Perturbation in The Shear Rate}
\label{sec:shear2}

The result in the previous section can be the starting point of a
perturbation theory in the shear rate in the case that $\tilde\gamma$
is non-zero, but small. From symmetry ($\gamma\rightarrow-\gamma$), it
follows that the Lyapunov exponent $\lambda^+$ is even in $\gamma$, so
to see the effect of the shear we need a second order perturbation
theory at least. We will sketch the solution formally in this section,
and leave the explicit calculations for the future.

We start with the eigenvalue equation in Eq.~(\ref{eq46}) to zeroth
order, i.e.,
\[
	{\bf L}^{0} A^{0} = \Lambda^{0} (\nu^*+s\bar\nu) A^{0} ,
\]
and write ${\bf L}^{0}={\bf L}^{0}_{0} +\tilde\gamma{\bf
L}^{0}_{1}+\tilde\gamma^2{\bf L}^{0}_{2} + {\cal O}(\tilde\gamma^3)$,
$\nu^* = \nu_{0}+ \tilde\gamma \nu^*_{1}+ \tilde\gamma^2 \nu^*_{2} +
{\cal O}(\tilde\gamma^3)$, $A^{0}=A^{0}_{0} + \tilde\gamma A^{0}_{1}+
\tilde\gamma^2 A^{0}_{2} + {\cal O}(\tilde\gamma^3)$ and
\begin{eqnarray}
	\Lambda^{0}(s) &=& \Lambda^{0}_{0}
		+ \tilde\gamma^2\Lambda^{0}_{2}(s)
		+ {\cal O}(\tilde\gamma^4)
\label{eq58}
\end{eqnarray}
where we used that $\Lambda^{0}$ is an even function of
$\tilde\gamma$.  For the same reason we may expand $\bar \nu$ as $\bar
\nu=\bar \nu_0+\tilde\gamma^2\bar \nu_2$ Substituting these
expressions into the eigenvalue problem and equating equal powers of
$\tilde\gamma$ one finds
\begin{eqnarray*}
A^{0}_{1} &=& {\cal R}\left[ \Lambda^{0}_{0}\nu^*_{1}
			-{\bf L}^{0}_{1}\right]A^{0}_{0}, \\
\Lambda^{0}_{2} &=& \frac{
	\left(A^{0}_{0},
	[
	{\bf L}^{0}_{2}-\Lambda^{0}_{0}(\nu_{2}^*+s\bar\nu_{2})]A^{0}_{0}
	+({\bf L}^{0}_{1} -\Lambda^{0}_{0}\nu^*_{1})A^{0}_{1}
	\right)
	}{
	(A^{0}_{0},(\nu_{0}+s\bar\nu_{0})  A^{0}_{0})
	},\\
A^{0}_{2} &=& {\cal R}
		(\Lambda^{0}_{0}\nu_{1}^*-{\bf L}^{0}_{1})A^{0}_{1}
\\&&
		+
	{\cal R}[\Lambda^{0}_{0}(\nu^*_{2}+\bar\nu_{2}s)
		 +\Lambda^{0}_{2}(\nu_{0}+
		 s\bar\nu_{0})-{\bf L}^{0}_{2}]
		A^{0}_{0},
\end{eqnarray*}
where ${\cal R}$ is the inverse of ${\bf
L}^{0}_{0}-\Lambda^{0}_{0}(\nu_{0}+s\bar\nu_{0})$, restricted to the
subspace orthogonal to $A^{0}_{0}$ (so $A^{0}_{1}$ and $A^{0}_{2}$ are
made unique by requiring orthogonality to $A^{0}_{0}$).

{}From Eqs.~(\ref{eq48}) and (\ref{eq58}), we obtain
$w^{0}(s)=w^{0}_{0}(s)+\tilde\gamma^2w^{0}_{2}(s)+ {\cal
O}(\tilde\gamma^4)$, where
\begin{eqnarray*}
	w^{0}_{0}(s)&=&-s/\ln\Lambda^{0}_{0}(s) \\
w^{0}_{2}(s)&=&s\Lambda^{0}_{2}/[\Lambda^{0}_{0}(s)\ln^2\Lambda^{0}_{0}(s)].
\end{eqnarray*}
The location of the minimum of the function $w^{0}(s)$ is shifted
by an amount of order $\tilde{\gamma}^2$, in fact
\begin{equation}
	s^{0} = s^{0}_{0} - \tilde\gamma^2 \frac{\frac{d
		w^{0}_{2}
		}{ds}(s^{0}_{0})
		}{\frac{d^2w^{0}_{0}}{ds^2}(s^{0}_{0})}
	+ {\cal O}(\tilde\gamma^4)
\label{eq59}
\end{equation}
as can be found from $(d/ds)w^{0}(s^{0})=0$ and 
$(d/ds)w^{0}_{0}(s^{0}_{0})=0$.
Because we are expanding around a minimum, this shift is not needed for
the value of the minimum of the function $w^{0}(s)$, i.e.,
\[
	w^{0}(s^{0}) = w^{0}_{0}(s^{0}_{0})
	+ \tilde\gamma^2
	w^{0}_{2}
	(s^{0}_{0})
	+{\cal O}(\tilde\gamma^4).
\]
Thus we can determine the shear corrections to the leading density term in
Eq.~(\ref{eq52}).

To calculate the density correction in Eq.~(\ref{eq52}) we do need the
shift of the minimum as it enters in the matrix elements.  This means
that one has to insert Eq.~(\ref{eq59}) and expand all relevant matrix
elements in powers of $\tilde\gamma$. All of these can be obtained
with the help of Eqs.~(\ref{eq17}), (\ref{eq19}), (\ref{eq20}),
(\ref{eq21}), (\ref{eq22}), (\ref{eq24}), (\ref{eq25}), (\ref{eq30}),
(\ref{eq44}) and (\ref{eq47}). In the end the evaluation will have to
be done numerically.

\section{Discussion}
\label{sec:conclusions}

In this paper we developed an analytic method for calculating the
largest Lyapunov exponent of a many body system, based on the
microscopic equations of motion.  To be specific we restricted
ourselves to uniform hard disk systems in two dimensions at low
densities, but not necessarily in equilibrium. As a particular case we
considered the uniformly sheared hard disk gas.

To obtain the largest Lyapunov exponent we derived a generalized
Boltzmann equation that describes the time evolution of a distribution
function of particle positions and velocities, together with deviation
vectors in tangent space.  At low densities the position deviations
turn out to be unimportant, to leading orders in the density. The
velocity deviation of a particle may be represented conveniently in
terms of the logarithms of its norm-- the so-called clock value $k$
(Eq.~(\ref{eq1})) -- and an additional unit vector, which in the end
plays no essential role.

As in the preceding papers \cite{myself,leiden,inszasz} the
generalized Boltzmann equation may be reinterpreted as describing the
propagation of a pulled front on the real line of possible clock
values. Therefore, by standard techniques the linearized version of
this equation may be used to obtain the asymptotic speed of
propagation of the front, which is directly proportional to the
largest Lyapunov exponent.

A remarkable property of the generalized Boltzmann equation is that
its natural density expansion does not proceed in powers of the
dimensionless density $\tilde{n}$, but rather in powers of
$\vartheta=1/|\ln\tilde{n}|$. To lowest order, this reproduces the
results of earlier work\cite{myself,inszasz}, which therefore finds a
firmer basis here. Here we also give values for the next order in
$\vartheta$ contribution to the largest Lyapunov exponent, and we give
explicit expressions for the first non-vanishing shear rate dependent
contribution to this exponent (quadratic in the shear rate) in the
uniformly sheared system.

In \cite{inszasz} we made comparisons between our results and the
results of numerical simulations of hard disk Lyapunov exponents by
Dellago and Posch\cite{Dellago}. Within the numerical accuracy good
agreement was found, but it turned out there are two complicating
factors. The first one is that the density should be very low to be in
the regime where a power series in $1/\ln(\tilde{n})$ can be expected
to
converge rapidly.  The second one is that there are large finite size
effects for the front speed $w$. One cannot (yet) reach the required
number of particles in simulations for these effects to become
negligible. The finite size effects can be estimated for
asymptotically large $N$ using front propagation techniques, and they
would scale as $1/ln^2(N)$. It is not known at which number of
particles this asymptotic result suffices, though.

To conclude we mention some possible extensions of this work and some
interesting problems that are still open for research. First of all
the expressions for the shear rate dependent contribution should be
worked out numerically.  We plan to do this on short notice. Then one
would like to go more general potentials, to general densities and to
three-dimensional systems as well as two-dimensional ones. We expect
that, as long as one stays at low densities, the generalization to
other simple (short ranged and spherically symmetic) potentials should
not pose any serious problems; one just has to use the generalized
Boltzmann equation that is appropriate for the potential under
consideration. Similarly we expect the generalization to three
dimensions just will cause some additional technical complications,
related to the fact that the directions of the velocity deviations
after a collision will depend on those before, unlike in two
dimensions. Generalizations to higher density will be much harder to
accomplish.  Already the fact that the natural expansion in density
proceeds in powers of $\vartheta$ rather than $\tilde{n}$ indicates
that a systematic expansion up to appreciable density will be
forbiddingly hard. On the other hand non-systematic approaches, such
as a generalized Enskog equation for hard disks or spheres, may give
good approximations, but this has not been explored yet.

One can also try to extend the methods developed here so as to
calculate additional Lyapunov exponents. For non-equilibrium systems
the most negative Lyapunov exponent is of primary interest, because
the sum of this and the largest Lyapunov exponent is directly related
to the shear viscosity through the conjugate pairing rule\cite{Evans}.
For the SLLOD equations with a Gaussian or ``constant-$\alpha$''
thermostat, there are deviations of this
rule\cite{Searles,PanjaVanZon,thesis}, but these are of higher order
in the shear rate, so that the connection with the {\em linear}
viscosity still stands.

Finally, at low densities, the methods developed here may probably be
combined with those of Refs.~\cite{KSentropy,KSentropy2} in order to
calculate the sum of all positive Lyapunov exponents, or the
KS-entropy, for non-equilibrium cases.  However, a theoretical
calculation of the full spectrum of Lyapunov exponents, which probably
requires similar techniques, remains a very challenging open problem.

\section*{Acknowledgements}

We dedicate this paper to our dear friend Bob Dorfman, whose enthusiasm
and hospitality contributed in a major way to its completion.

We acknowledge support by the Mathematical physics program of FOM and
NWO/GBE and by the National Fund for Scientific Research (F.\ N.\ R.\
S.\ Belgium).

\begin{figure}[h]
	\caption{Velocity profile in a gas under shear}
\label{fig:shear}
\end{figure}

\pagebreak

\begin{figure}[h]
\begin{center}
      \epsfig{figure=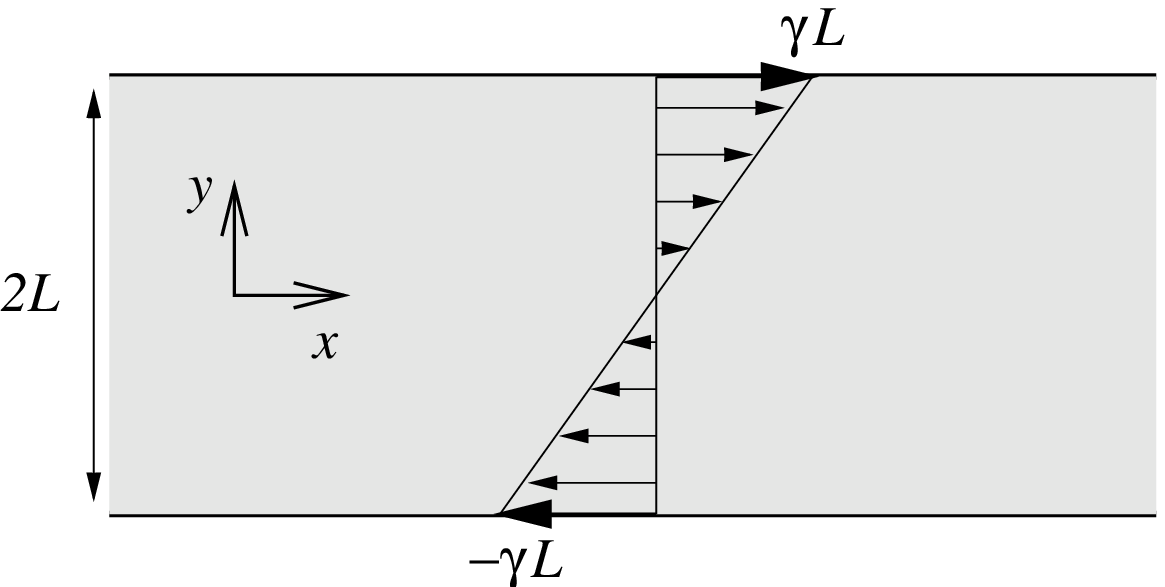}
\end{center}
\end{figure}

\end{document}